\documentclass[a4paper,reqno]{amsart}
\usepackage{geometry}
\usepackage{fullpage}
\usepackage[latin1]{inputenc}
\usepackage[T1]{fontenc}
\usepackage{amsfonts}
\usepackage{color}
\usepackage{epsfig}
\usepackage{epstopdf}
\usepackage{subfigure}
\usepackage{bm}
\usepackage{bbm}
\usepackage{bbold}
\usepackage{amssymb}
\usepackage{amsmath}
\usepackage{amsthm}
\usepackage[foot]{amsaddr}
\usepackage{graphicx}
\usepackage{subfigure}
\usepackage[dvipsnames]{xcolor}
\usepackage{afterpage}
\usepackage{booktabs}
\usepackage{siunitx}

\usepackage{booktabs}
\usepackage{fix-cm}

\usepackage[colorlinks=true, linkcolor=blue, citecolor=cyan, urlcolor=green]{hyperref}

\graphicspath{{plots/}}

\numberwithin{equation}{section}

\definecolor{light}{gray}{.9}

\def\be{\begin{equation}}
\def\ee{\end{equation}}
\def\bea{\begin{eqnarray}}
\def\eea{\end{eqnarray}}

\def\E{\mathbb{E}}

\newcommand{\ie}{\textit{i.e. }}

\newcommand{\nocontentsline}[3]{}
\newcommand{\tocless}[2]{\bgroup\let\addcontentsline=\nocontentsline#1{#2}\egroup}

\DeclareMathSymbol{\leqslant}{\mathalpha}{AMSa}{"36} 
\DeclareMathSymbol{\geqslant}{\mathalpha}{AMSa}{"3E} 
\DeclareMathSymbol{\eset}{\mathalpha}{AMSb}{"3F}     
\renewcommand{\leq}{\;\leqslant\;}                   
\renewcommand{\geq}{\;\geqslant\;}                   

\def\P{\mathbb{P}}
\def\E{\mathbb{E}}

\title{When panic makes you blind:\\ a chaotic route to systemic risk}
\date{\today}

\author[1]{Piero Mazzarisi$^{1,2}$}
\address[1]{Scuola Normale Superiore, Pisa, Italy}
\email[1]{piero.mazzarisi@sns.it}
\email[1]{piero.mazzarisi@unibo.it}
\author[2]{Fabrizio Lillo$^2$} 
\address[2]{Department of Mathematics, University of Bologna, Italy}
\email[2]{fabrizio.lillo@unibo.it}
\author[3]{Stefano Marmi$^1$}
\email[3]{stefano.marmi@sns.it}


\begin{document}
\maketitle

\begin{abstract}
We present an analytical model to study the role of expectation feedbacks and overlapping portfolios on systemic stability of financial systems. Building on \cite{corsi2013when}, we model a set of financial institutions having Value at Risk capital requirements and investing in a portfolio of risky assets, whose prices evolve stochastically in time and are endogenously driven by the trading decisions of financial institutions. Assuming that they use adaptive expectations of risk, we show that the evolution of the system is described by a slow-fast random dynamical system, which can be studied analytically in some regimes.  The model shows how the risk expectations play a central role in determining the systemic stability of the financial system and how wrong risk expectations may create panic-induced reduction or over-optimistic expansion of balance sheets. Specifically, when investors are myopic in estimating the risk, the fixed point equilibrium of the system breaks into leverage cycles and financial variables display a bifurcation cascade eventually leading to chaos. We discuss the role of financial policy and the effects of some market frictions, as the cost of diversification and financial transaction taxes, in determining the stability of the system in the presence of adaptive expectations of risk.  
\end{abstract}

\section{Introduction}
Borrowing is an essential aspect of the business of financial institutions operating in financial markets because of the possibility of leveraging the investment returns by buying on margin. The recent empirical literature \cite{adrian2010liquidity} captured perfectly this aspect characterizing the leverage management of the largest investors operating in the market. But financial leverage is directly related to risk. As several recent papers suggest \cite{brunnermeier2009market,geanakoplos2010leverage,aymanns2015dynamics,corsi2013when}, financial leverage is probably the most important engine in driving endogenously the booms and the busts of asset prices in the market. 

Regulators try to limit the exacerbated use of leverage by imposing some constraints to financial institutions in order to make the financial system more robust and resilient to shocks. Value-at-Risk (VaR) constraint is probably the most popular one, but other more sophisticated ones have been proposed in recent years by the successive Basel regulations. All the regulators' constraints require an estimation of the riskiness of the investments as well as of the dependencies between extreme events of financial returns.  Therefore both the capital requirement constraint and the risk/dependency expectations play a crucial role in determining the systemic stability of financial markets. 

On one hand, as documented by many papers (see for example \cite{adrian2010liquidity}),  VaR capital requirements, as other risk constraints, can induce a perverse demand function: in order to target leverage,  a  financial institution will sell more assets if their price drops and viceversa when their price rises. Thus, a marked-to-market and VaR constrained financial institution will have a positive feedback effect on the prices of the assets in its portfolio. In a market where many financial institutions are forced to follow similar risk management rules, the coordinated rebalancing of portfolios composed by illiquid assets creates a feedback effect which depends on the degree of leverage and diversification. Thus, while diversification of investment should reduce portfolio risk, a significant overlap (i.e. similarity) of portfolios of many financial institutions can instead destabilize the market and increase its susceptibility to price shocks, as documented by fire sales spillovers \cite{greenwood2015vulnerable}.

On the other hand, the implementation of any capital requirement depends on the expectations financial institutions have on the risk of the assets in the portfolio and on their statistical dependence. For this reason there is a vast literature on the estimation of risk and dependences \cite{tsay2005analysis,hommes2007learning,heemeijer2009price,bao2013learning}, many of them based on the recent history of prices in a time window of the recent past. The choice of the length of the estimation window is critical, since  there is a tradeoff between choosing a long estimation window to improve statistical significance and preferring a short window in order to capture a more timely measure of risk. In period of financial turbulence, where non stationary effects are more likely, investors might prefer to use short estimation windows. Since trading decisions of financial institutions in order to comply with capital requirements depend on the estimated risk, in the presence of illiquid assets the length of the risk estimation window can impact the dynamical properties of prices. This creates a second feedback effect in addition to the one, described above, due to target leveraging.

In this paper we present an analytical model of the financial system where both feedbacks mechanisms are present. Building on \cite{corsi2013when}, a set of financial institutions (banks) investing in a portfolio of risky and illiquid assets follow a target leveraging strategy to satisfy Value at Risk capital requirements. The estimations of risk of the investment assets, and as a consequence the leverage, are periodically updated and banks use a backward-looking expectation scheme which considers price returns in a past time window to build estimates. The two feedback mechanisms are coupled by the price dynamics, which on one side is used to mark-to-market the portfolio and to estimate risk and correlations, and on the other one is endogenously affected by the trading activity of financial institutions.    

Interestingly the two feedback mechanisms described above act on different time scales. In our model the time scale of leverage targeting is shorter than the time scale over which financial institutions update their risk expectations. This separation of time scales is crucial in our modeling. Since the slow variables, associated with updates of risk expectations, evolve in time as a function of averages over the fast variables, associated with leverage targeting, our model can be casted as a discrete time slow-fast dynamical system\footnote{The mathematical framework for slow-fast systems at continuous time is studied in \cite{kuehn2011mathematical}. A review about random dynamical systems is \cite{bhattacharya2003random}. The averaging principle can be considered as an extension of the ergodic theorem: the case of interest for our purposes, where a deterministic dynamical system is coupled with a random one, is discussed for example in \cite{dolgopyat2004evolution,dolgopyat2005introduction,kifer2014nonconventional}.}. The ratio between the two time scales is the key parameter determining the type of mathematical modeling. We show that when this ratio tends to infinity, i.e. financial institutions are continuously marked-to-market, the dynamics is described by a deterministic map. The window used to form expectations of risk plays a central role in determining systemic stability and leverage cycles appear when investors become more myopic relative to past history of asset prices, \ie this memory becomes smaller than a given threshold. Our model predicts that the deterministic dynamics of the financial system becomes chaotic when the memory decreases further and goes below a second smaller threshold. When the ratio between the two time scales is finite a random slow-fast dynamical system describes the system. Even if mathematically this is harder to study, because of the joint  chaotic and stochastic dynamics, we show by analytical arguments and numerical simulations  that the main dynamical characteristics remain unchanged.

We are therefore able to characterize the possible dynamical outcomes for the considered financial system as a function of the  memory window used to form expectations, the tail parameter of the Value at Risk, the number of asset classes available for investment,  the ratio between the two time scales (related to the presence of market frictions), and a parameter determining the level of financial innovation. We show how the breaking of the fixed point equilibrium for the financial system occurs via a period-doubling bifurcation when any of these parameters are varied and how the dynamics of the financial system may be intrinsically chaotic in certain parameter regions. Each of these parameters can at least in part controlled by regulators, thus our model is able to provide policy recommendation for enhancing financial stability, as discussed at the end of the paper. 

\paragraph{\bfseries{Related literature.}}
Our paper aims to combine several streams of literature: (i) the analysis of portfolio rebalancing induced by the mark-to-market accounting rules and VaR constraint \cite{adrian2010liquidity,adrian2014procyclical}; (ii) the investigations on the impact of the imposition of capital requirements on the behaviour of financial institutions and their possible procyclical effects \cite{danielsson2004impact,danielsson2012endogenous,tasca2013market,adrian2014procyclical,corsi2013when}; in particular, we generalize the model of Corsi et al. by studying how the procyclical effects on asset prices and risk expectations influence the portfolio decisions about leverage and diversification; (iii) the literature on distressed selling and its impact on the balance sheets of the financial institutions because of the overlapping among portfolios \cite{cont2013running,caccioli2012stability}; (iv) the research on the role of expectation feedbacks in financial systems \cite{hommes1994dynamics,hommes2000cobweb,farmer2012complex,hommes2013behavioral}; in particular, our paper focuses on the role of risk expectations which are formed by using statistical models of past observations of investment prices; (v) finally our paper contributes to literature on the application of dynamical systems theory to the problem of systemic risk in financial markets \cite{douady2012financial,castellacci2014modeling} and, specifically, to the study of the dynamics of leverage cycles and its relation with the financial regulation \cite{brunnermeier2009market,geanakoplos2010leverage,poledna2014leverage,aymanns2015dynamics,aymanns2016taming,halling2016leverage}.

\paragraph{\bfseries{Outline.}}
The remainder of this paper is organized as follows. In Section \ref{model} we briefly review the model of \cite{corsi2013when} which represents the background of our study. Then, we introduce the model with backward-looking risk expectations. In Section \ref{corsiresult} we briefly summarize the results of the paper \cite{corsi2013when} which represents the limit case of our model without backward-looking expectations. In Section \ref{ds} we analyze the dynamical properties of the model and its policy implications for financial markets in the limit in which the model is fully analytical. This limit corresponds to study the deterministic skeleton of the slow-fast random dynamical system. In Section \ref{fluctuations}, we present an analytical argument to extend the obtained results also in the random framework and we give more insights into the model via numerical simulations and discuss the role of taxation for financial systemic risk. Section \ref{conclusions} contains some conclusions.

\section{Financial system with expectation feedbacks}\label{model}
We model a financial market consisting of leveraged financial institutions investing in some risky assets. Financial institutions, here called banks, have capital requirements in the form of VaR constraint and face costs of diversification in forming a portfolio. We focus on the study of systemic risk in financial markets from the point of view of indirect contagion of risk. In presence of investments' illiquidity, losses and distressed selling propagate indirectly among financial institutions by common investments in their portfolios. 

The adopted framework is the one introduced in \cite{corsi2013when}, where authors model a bipartite network of investment assets and banks. To form a portfolio each bank solves an optimization problem to determine simultaneously the optimal values of leverage and diversification. Determining the optimal level for the financial leverage is related to the fulfilment of the VaR constraint. Hence, portfolio decisions about leverage and diversification depend on the expectations of risk by banks.

\cite{corsi2013when} are agnostic on the process of expectation formation, since expectations of returns and risks are described by exogenous parameters. As a consequence, there is no dynamics associated with the evolving banks' expectations.

Here we explicitly introduce a process for the formation of risk expectations\footnote{Expected asset returns remain exogenous to the price dynamics. This is consistent with the view of  financial institutions as fundamental traders.} where banks estimate the risk of their portfolio in the holding period from past price movements \cite{hommes2009handbook}. In our model, we introduce a simple process for the backward-looking expectations which are characterized by one parameter related to the memory of expectations. The mechanism of expectation feedbacks induces a dynamical component in the portfolio decisions: depending on the evolution of market prices, the perception of risk by banks changes and accordingly also the portfolio decisions about leverage and diversification. Since asset prices evolve stochastically in time, the market value of the portfolio changes. By assuming that the liability size remains unchanged in the meanwhile, the exogenous variation of the leverage occurs in the inverse direction of the portfolio, \ie if the prices increase, the financial leverage decreases. As empirically found \cite{adrian2010liquidity,adrian2014procyclical}, banks manage actively their balance sheet by rebalancing periodically the portfolio in order to have the leverage equal to its optimal level or target value. The time scale associated with the portfolio rebalancing is the time spent by banks to bring the leverage back to its target value. The frequency of balance sheet adjustments determines how well the bank is marked-to-market in the portfolio holding period. 

In our model, we define the time scale of the dynamics for portfolio decisions as the unit time scale and  $n$ is the number of times a bank rebalances its portfolios for leverage targeting in the unitary holding period. Hence, the time scale associated with the portfolio rebalancing is $1/n$.

In presence of finite investments' liquidity, the trading of assets drives endogenously the dynamics of the price. Since the price history is used to form expectations of risk, a feedback coupling portfolio decisions and price dynamics is created. 

Below, we briefly review the model of \cite{corsi2013when}. Then, we will describe our model.

\subsection{Baseline model of \cite{corsi2013when}}

The system is composed by $N$ financial institutions (banks) creating portfolios by investing in some of the $M$ available risky assets. In the following, we adopt the convention of labeling risky investments as $i,j,k,\ldots$ and banks as $a,b,c,\ldots$.  All financial institutions are assumed to be equivalent, that is they have the same initial equity capital and the same capital requirements, they solve the same portfolio problem and they use the same scheme in forming risk expectations. This is of course a simplifying assumption in order to achieve analytical tractability. However, \cite{corsi2013when} show how heterogeneity does not affect crucially the results of the model.

\begin{figure}[t]
\centering
{\includegraphics[width=0.65\textwidth]{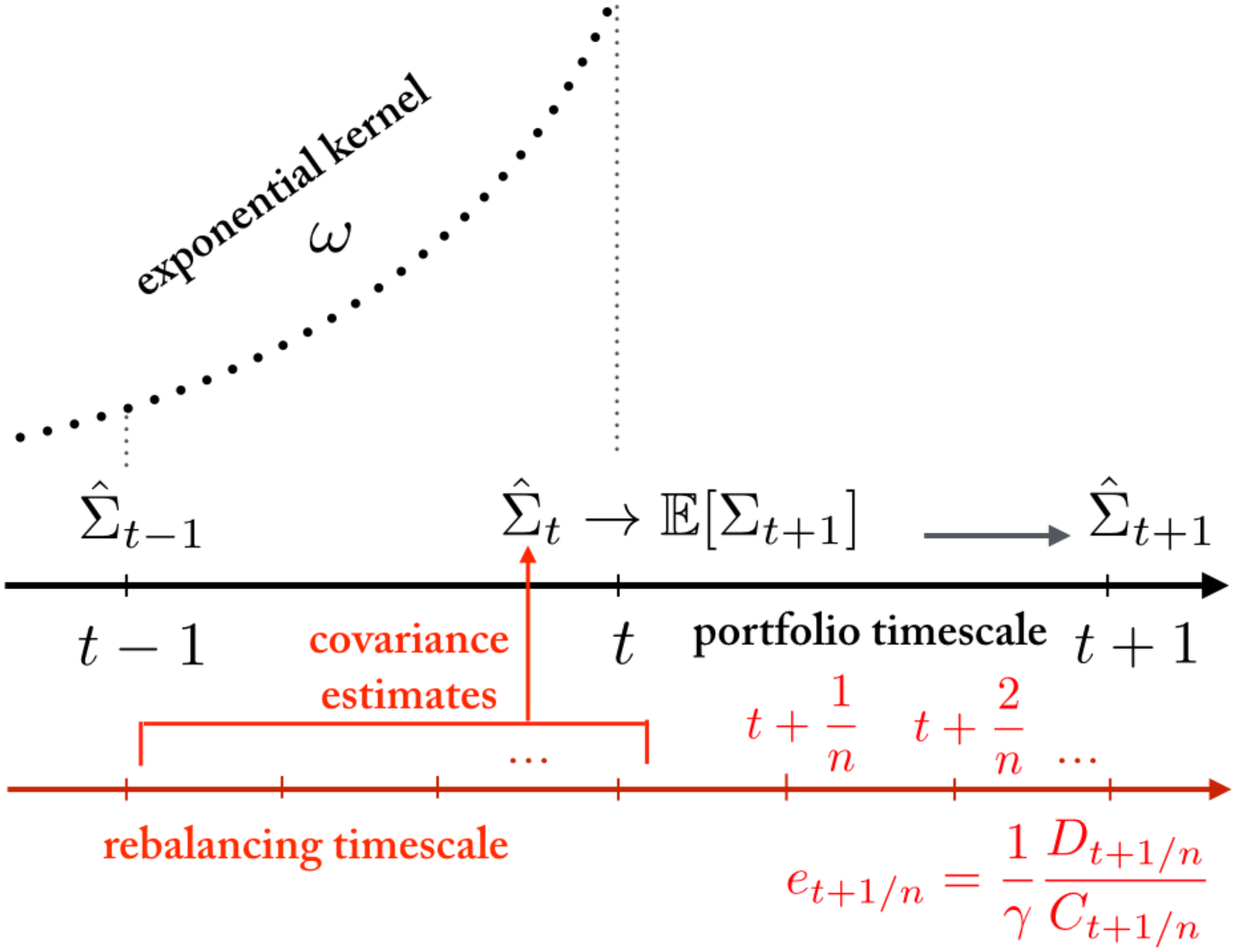}}
\caption{Timing of the slow-fast dynamical model describing the financial system.}\label{timescales}
\end{figure}

We assume that banks update their expectations of risk at time intervals of unitary length and accordingly they make new decisions about leverage and portfolio diversification (see Fig. \ref{timescales}). In the time interval $(t,t+1]$, ($t \in \mathbb{Z}$), banks rebalance their portfolio to target the leverage, but without changing the risk expectations. The rebalancing takes place in $n$ time subintervals within $(t,t+1]$, \ie at $\{t+\frac{1}{n},t+\frac{2}{n},t+\frac{3}{n},...,t+\frac{n}{n}\equiv t+1\}$. $\frac{1}{n}$ is the timescale associated with the {\it portfolio rebalancing} (the red axis in Figure \ref{timescales}) whereas the unit value is the timescale associated with the {\it portfolio decisions} (the black axis in Figure \ref{timescales}). Clearly, $n\in\mathbb{N}\setminus\{0\}$. As clarified below, \cite{corsi2013when} considered the case of removing the endogenous process of formation of risk expectations, \ie portfolio decisions are taken without consideration on the price dynamics. This is equivalent to optimize the portfolio only once with exogenous risk expectations and the process of portfolio rebalancing follows without changing the value of the optimal leverage.

\paragraph{\bfseries{Portfolio decisions.}} 
The equity of each financial institution at time $t$ is given by $E_t=A_t-L_t$, where $L_t$ ($A_t$) represents the liabilities (assets)  of the bank at time $t$. For simplicity banks do not face funding restrictions and they can decide to increase or to reduce its liabilities according to their needs as long as they fulfil the VaR constraint. Finally $r_L$ is the per dollar average interest expense on the liability side and the financial leverage is $\lambda_t=\frac{A_t}{E_t}$.

\cite{corsi2013when} considered a simplified setting where each bank determines the optimal value for leverage and diversification. Since all the risky investments are ex-ante statistically equivalent, financial institutions adopt a simple investment strategy consisting in forming an equally weighted portfolio by randomly selecting $m$ risky investments from the whole collection of $M$ available investment assets. Hence, banks have to find the optimal number of investment assets in the presence of diversification costs $\bar{c}$, see \cite{constantinides1986capital}. The costs of diversification represent all the informational and infrastructural costs  preventing each institution to achieve full diversification of its portfolio. 

Banks select also the optimal value for leverage $\lambda_t$, being bounded by the VaR constraint $VaR\times A_t\leq E_t$. By assuming a functional shape of the return distribution, $VaR$ is the $\alpha$-multiple of the expected holding period volatility of the portfolio $\sigma^p$, \ie $\alpha\sigma^p$. In the numerical experiments below, the VaR quantile associated with a $P_{VaR}=5\%$ is given by $\alpha=1.64$ if we assume normally distributed portfolio returns.

Since investments are equivalent, $\mu$ is the expected return in the holding period and then the net interest margin of the financial institutions is $\mu-r_L$.  Finally, financial institutions correctly perceive that each risky investment entails both an idiosyncratic (diversifiable) risk component and a systematic (undiversifiable) risk component, \ie the expected variance of an investment $i$ is $\sigma^2_{i,t}=\Sigma_{u,t} +\Sigma_{d,t}$, where the first (second) term represents the expected variance of the systematic factor (idiosynchratic noise). 

In the following we use the hat, e.g. $\widehat{\Sigma}_{u,t}$, to indicate an \emph{estimation} of the variance formed by using past observations. On the contrary, unlabeled symbols indicate \emph{expectation}; for example, $\Sigma_{u,t}$ is the expectation at time $t$ of the systematic risk at time $t+1$.

The expected variance of a portfolio of $m_t$ assets is 
\begin{equation}\label{portfolio_risk}
\sigma^2_{p,t}=\frac{\Sigma_{d,t}}{m_t}+\Sigma_{u,t}.
\end{equation}
As shown in \cite{corsi2013when}, the portfolio optimization problem is
\begin{equation}\label{portfolio_opt}
\max_{\lambda_t,m_t} \lambda_t(\mu-r_L)-cm_t\:\:\:\mbox{ s.t. }\:\:\:\alpha{\sigma_{p,t}}\lambda_t\leq 1
\end{equation}
where $c$ measures the diversification cost for each investment. The solution in implicit form is 
\begin{equation}\label{eqm}
m_t=\lambda_t\:\sqrt{\Sigma_{d,t}}\sqrt{\frac{\alpha}{2c}\frac{\mu-r_L}{{\sigma_{p,t}}}}
\end{equation}
\begin{equation}\label{eql}
\lambda_t=\frac{1}{\alpha\sigma_{p,t}}.
\end{equation}

Given $\Sigma_{d,t}$ and $\Sigma_{u,t}$, there is only one independent variable in the system of Equations \ref{eqm} and \ref{eql}. Indeed, it is
\begin{equation}\label{mdependent}
m_t\equiv m_t(\lambda_t; \Sigma_{d,t},\Sigma_{u,t})=\frac{\Sigma_{d,t}}{\frac{1}{\alpha^2\lambda_t^2}-\Sigma_{u,t}}
\end{equation}
where $\lambda_t$ is the only positive real solution\footnote{It can be shown that Eq. \ref{lambdamap} has only one positive real solution for $\lambda_t\in\mathbb{R}^+$ in the space of feasible parameters, \ie $\alpha>0$, $c\in[0,1]$, $\mu-r_L>0$, $\Sigma_{d,t},\Sigma_{u,t}>0$.} of the following quartic equation,
\begin{equation}\label{lambdamap}
\Sigma_{d,t}^{\frac{1}{3}}\left(\alpha^{\frac{2}{3}}\left(\frac{\mu-r_L}{2c}\right)^{\frac{1}{3}}\lambda_t\right)^{-1}-\left(\frac{1}{\alpha^2\lambda_t^2}-\Sigma_{u,t}\right)^{\frac{2}{3}}=0.
\end{equation}

\paragraph{\bfseries{Portfolio rebalancing.} }

As empirically shown by \cite{adrian2010liquidity}, financial institutions adopting target leverage adjust their assets and liabilities rather than raising or redistributing equity capital. This means that equity changes over time as a consequence of the bank profits and losses, but it is not actively managed. 

Similarly to \cite{corsi2013when}, we model the dynamics of the return of the risky investment $i$ in the $n$ time steps of length $1/n$ during the interval $(t,t+1]$  as the sum of two components:
\begin{equation}\label{return}
r_{i,t+k/n}=\varepsilon_{i,t+k/n}+e_{i,t+(k-1)/n}, ~~~~~~~~~~k=1,2,...,n.
\end{equation}
Note that here the time scale is the fast one, thus we use a fractional time labeling to indicate the times between $t$ and $t+1$ when portfolio rebalancing occurs.
The exogenous component $\varepsilon_{i,t+k/n}=\mu_1+f_{t+k/n}+\epsilon_{i,t+k/n}$ is the sum of a drift term representing the expected return  plus a systematic market factor $f_{t+k/n}$ common to all risky investments, plus a noise term $\epsilon_{i,t+k/n}$ representing idiosyncratic innovation. Without loss of generality, both the noise term and the systematic factor are Gaussian distributed, \ie $\epsilon_{i,t+k/n}\sim\mathcal{N}(0,\sigma_\epsilon^2)$ and $f_{t+k/n}\sim\mathcal{N}(0,\sigma_f^2)$ $\forall k=1,2,...,n$.

The endogenous component $e_{i,t+(k-1)/n}$ depends on the price impact of the demand for the risky investment $i$ arising from portfolio rebalancing. Given a target leverage, when asset prices increase\footnote{When the asset prices decrease, the sign is simply reverted.}, both asset value and equity increase, since liabilities remain constant, and as a consequence the leverage  decreases\footnote{E.g. if the profit is equal to $\delta A$, we have $\frac{A+\delta A}{E+\delta A}<\frac{A}{E}$.}. Thus, in order to keep leverage equal to the target value, banks manage their balance sheet by increasing the liabilities and using the borrowed money for the purchase of new assets\footnote{According to the assumption of statistically equivalence of risky investments, the increment of asset size is equally distributed over assets in the portfolio \cite{greenwood2015vulnerable}.}. In the presence of asset illiquidity, buying or selling assets for leverage targeting will move their prices. 

In mathematical terms, given at the generic fractional time $s=t+k/n,\:\:k=1,2,...,n$  the desired asset size for a generic bank $a$, $A_{a,s}^*=\lambda_t E_{a,s}$, the bank $a$ rebalances the portfolio by trading a quantity which is the difference between the desired asset size and the current one, $\Delta R_{a,s}=A_{a,s}^*-A_{a,s}$, which is given by
\begin{equation}\label{rebalancing}
\begin{split}
\Delta R_{a,s} &=  A^*_{a,s}-A_{a,s}=\lambda_t E_{a,s}-A_{a,s-1/n}^*(1+r_{a,s}^p)=\\
& = \lambda_t (E_{a,s-1/n}+r_{a,s}^pA_{a,s-1/n}^*)-A_{a,s-1/n}^*(1+r_{a,s}^p)=\\
& = (\lambda_t-1)A_{a,s-1/n}^* r_{a,s}^p
\end{split}
\end{equation}
where $r_{a,s}^p$ is the portfolio return at time $s$ for the bank $a$. Eq. \ref{rebalancing} shows that any profit or loss from investments in the portfolio ($A_{a,s-1/n}^* r_{a,s}^p$) will result in a change in the asset size, amplified by financial leverage.

The impact of the portfolio rebalancing by all banks at time $s$ will affect asset prices at time $s+1/n$. The total demand of the risky investment $i$ at time $s$ will be the sum of the individual demand of the banks who picked investment $i$ in their portfolio, 
\begin{equation}\label{demand}
D_{i,s}=\sum_{a=1}^NI_{i\in a}\frac{1}{m_t}\Delta R_{a,s}.
\end{equation}
where $I_{i\in a}$ is the indicator function which takes value one when investment $i$ is in the portfolio of institution $a$ and zero otherwise. \cite{corsi2013when} show that this quantity can be written as  
$$
D_{i,s}=(\lambda_t-1)\frac{A_{s-1/n}^*}{m_t}\frac{N}{M}\left(r_{i,s}+\frac{m_t-1}{M-1}\sum_{k\neq i}r_{k,s}\right).
$$
By assuming that price impact is linear in the traded volumes,  \cite{corsi2013when} model  the endogenous component as
\begin{equation}\label{endo1}
e_{i,s}=\frac{1}{\gamma}\frac{D_{i,s}}{C_{i,s}}
\end{equation}
where $\gamma$ is a parameter measuring the investment liquidity, $C_{i,s}\equiv\sum_{a=1}^NI_{i\in a}\frac{A^*_{a,s-1/n}}{m_t}$ is a proxy for capitalization of investment $i$ at time $s$. Coherently with the assumption of statistically equivalence for risky investments, we assume that each investment is characterized by the same liquidity parameter $\gamma$.  \cite{corsi2013when} obtain the following Vector Autoregressive (VAR) dynamics for the vector of endogenous components
\begin{equation}\label{endo2}
\bm{e}_s=\bm{\Phi}_t\bm{r}_s=\bm{\Phi}_t(\bm{e}_{s-1/n}+\bm{\varepsilon}_s),\:\:\:\:s=t+k/n,\:\:k=1,2,...,n
\end{equation}
where 
\begin{equation}\label{PHImatrix}
\bm{\Phi}_t=\frac{(\lambda_t-1)}{\gamma}\left[
\begin{array}{cccc}
\frac{1}{m_t}&\frac{1}{m_t}\frac{m_t-1}{M-1}&\ldots& \frac{1}{m_t}\frac{m_t-1}{M-1}\\
 \frac{1}{m_t}\frac{m_t-1}{M-1}& \frac{1}{m_t}&\ldots& \frac{1}{m_t}\frac{m_t-1}{M-1}\\
\vdots& &\ddots&\vdots\\
 \frac{1}{m_t}\frac{m_t-1}{M-1}& \frac{1}{m_t}\frac{m_t-1}{M-1}&\ldots& \frac{1}{m_t}
 \end{array}\right].
\end{equation}
Notice that $\bm{\Phi}_t\equiv\bm{\Phi}(\lambda_t,\Sigma_{d,t},\Sigma_{u,t})$ because $m_t\equiv m_t(\lambda_t,\Sigma_{d,t},\Sigma_{u,t})$ according to Eq. \ref{mdependent}. 

The VAR(1) process in Eq. \ref{endo2} determines the return process in Eq. \ref{return} and completely defines the fast dynamics of our model between $t$ and $t+1$.


\subsection{\bfseries{Formation of risk expectations.}}\label{formationProcess} 
Here we generalize the model of \cite{corsi2013when} by specifying how banks form risk expectations in Eq. \ref{lambdamap} to take portfolio decisions. Several empirical and experimental studies, for example \cite{hommes2009handbook,bao2013learning}, have shown that financial agents use statistical models of past observations of the price to forecast the future, the so-called backward-looking expectations. Here we introduce a similar expectation scheme which exploits past observed prices in forecasting future risks. 

\paragraph{{\bfseries Risk estimation}}At each portfolio decision time $t$, banks estimate the covariance matrix of risky investments between $t-1$ and $t$, that is
\begin{equation}\label{covest}
\widehat{\bm{\Sigma}}_t=\left(
\begin{array}{ccc}
\widehat{Var}[R_{i,t}] & \widehat{Cov}[R_{i,t},R_{j,t}] & \cdots \\
\widehat{Cov}[R_{i,t},R_{j,t}] & \ddots & \vdots \\
\vdots & \cdots & \cdots \\
\end{array}
\right)
\end{equation}
where $\widehat{Var}[R_{i,t}]$ and $\widehat{Cov}[R_{i,t},R_{j,t}]$ with $i\neq j$ are the maximum likelihood estimators of the variance and covariance of assets' returns aggregated at the time scale of the portfolio decisions, \ie $R_{i,t}\equiv\sum_{k=1}^n r_{i,t-1+k/n}$\footnote{For notational simplicity, we are considering that the asset returns are centered around the mean.}. We assume that banks correctly perceive the return dynamics as evolving according to the vector autoregressive process VAR(1), see Eqs. \ref{return}, \ref{endo2} and \ref{PHImatrix}, as well as the statistical equivalence of asset investments. As a consequence, symmetric conditions can be imposed on $\widehat{\bm{\Sigma}}_t$, namely the diagonal components of the covariance matrix are equal to each other and the same for the off-diagonal components. It is
\begin{equation}\label{decompositionSigma}
\widehat{\bm{\Sigma}}_t=\widehat{\Sigma}_{d,t}\mathbb{I}+\widehat{\Sigma}_{u,t}\bm{1}
\end{equation}
where $\mathbb{I}$ is the identity matrix, $\bm{1}$ is the $M\times M$ matrix whose entries are equal to one and $\widehat{\Sigma}_{d,t}
=\widehat{Var}[R_{i,t}]-\widehat{Cov}[R_{i,t},R_{j,t}]$ and $\widehat{\Sigma}_{u,t}
=\widehat{Cov}[R_{i,t},R_{j,t}]$ $\forall i,j=1,...,M,\:\:\: i\neq j$.

We do not exploit explicitly the maximum likelihood estimators of the variance and covariance of returns to obtain the results in the next Section. However, we show in Appendix \ref{covarianceAppendix} how to obtain the explicit formulas for completeness. Here, let us stress that $\widehat{\Sigma}_{d,t}$ and $\widehat{\Sigma}_{u,t}$ are only functions of the fast variables $\bm{r}_{t-1+k/n},\:\:k=0,1,2,...,n$, \ie
$$
\begin{cases}
\widehat{\Sigma}_{d,t}\equiv\widehat{\Sigma}_{d,t}(\{r_{i,t-1+k/n}\}_{k=0,1,...,n}^{i=1,...,M})\\
\widehat{\Sigma}_{u,t}\equiv\widehat{\Sigma}_{u,t}(\{r_{i,t-1+k/n}\}_{k=0,1,...,n}^{i=1,...,M}).
\end{cases}
$$

\paragraph{{\bfseries Risk expectations}} Once banks have estimated the covariance matrix of risky investments, they form risk expectations of the two independent quantities of the covariance matrix, \ie $\Sigma_{d,t}$ and $\Sigma_{u,t}$. We assume the banks expectation at time $t$ of an element of the covariance matrix is a weighted sum of the previously adopted expectation for the same element and the current estimation, that is
\begin{equation}\label{expectations}
\begin{cases}
& \Sigma_{d,t}^\omega=\omega\Sigma_{d,t-1}^\omega+(1-\omega)\widehat{\Sigma}_{d,t}\\
& \Sigma_{u,t}^\omega=\omega\Sigma_{u,t-1}^\omega+(1-\omega)\widehat{\Sigma}_{u,t}
\end{cases}
\end{equation}
where $\omega\in[0,1]$ is the memory parameter of the expectation scheme. Eq. \ref{expectations} is also known as adaptive expectations \cite{hommes2013behavioral}. By iterating Eq. \ref{expectations}, we can write adaptive expectations as an Exponentially Weighted Moving Average (EWMA) of past observed risks, that is
\begin{equation}\label{expectations2}
\Sigma_{\cdot,t}^\omega=(1-\omega)\sum_{k=0}^{\infty}\omega^k\widehat{\Sigma}_{\cdot,t-k}=(1-\omega)\sum_{k=0}^{\infty}e^{-\frac{k}{\tau}}\widehat{\Sigma}_{\cdot,t-k}
\end{equation}
where $\tau=(\ln\frac{1}{\omega})^{-1}$ corresponds to the `effective' memory of adaptive expectations. The limit $\omega\rightarrow 0^+$ corresponds to naive expectations of risk, \ie no memory ($\tau\rightarrow0$). By increasing $\omega$, the effective memory $\tau$ of the expectation scheme increases.

\subsection{Expectation feedback system}\label{asymptotics}
By summarizing the building blocks of our model, the dynamics of the financial system is described by a slow-fast random dynamical system at discrete time which is specified by the following equations,

\begin{equation}\label{slow_d}
\begin{cases}
\left(\Sigma_{d,t}^\omega\right)^{\frac{1}{3}}\left(\alpha^{\frac{2}{3}}\left(\frac{\mu-r_L}{2c}\right)^{\frac{1}{3}}\lambda_t\right)^{-1}-\left(\frac{1}{\alpha^2\lambda_t^2}-\Sigma_{u,t}^\omega\right)^{\frac{2}{3}}=0 \\
\Sigma_{d,t}^\omega=\omega\Sigma_{d,t-1}^\omega+(1-\omega)\widehat{\Sigma}_{d,t}\\
\Sigma_{u,t}^\omega=\omega\Sigma_{u,t-1}^\omega+(1-\omega)\widehat{\Sigma}_{u,t}
\end{cases}
\end{equation}
\begin{equation}\label{fast_d}
\bm{r}_s=\bm{\varepsilon}_s+\bm{\Phi}(\lambda_{t-1},\Sigma_{d,t-1}^\omega,\Sigma_{u,t-1}^\omega)\:\bm{r}_{s-1/n}\:\:\: s=t-1+k/n,\:\:k=1,2,...,n
\end{equation}
where $\widehat{\Sigma}_{d,t}$ and $\widehat{\Sigma}_{u,t}$ are the maximum likelihood estimators of the (diversifiable) variance and covariance of returns aggregated at the time scale of portfolio decisions and can be obtained as explained in Appendix \ref{covarianceAppendix}, see Eq. \ref{SigmaHat}.

Eq. \ref{slow_d} describes the dynamics of banks' portfolio decisions, \ie the slow component of the dynamics described by the slow variables $\lambda_t$, $\Sigma_{d,t}^\omega$, and $\Sigma_{u,t}^\omega$. 
Eq. \ref{fast_d} describes the price evolution of the risky investments, \ie the fast component of the dynamics described by the fast variables $r_{i,s}$.

The model must satisfy a stationarity condition, namely covariance stationarity of the autoregressive process, which is $\lambda<\gamma+1$\footnote{The VAR(1) process in Eq. \ref{fast_d} is covariance stationary when the largest eigenvalue of the matrix $\bm{\Phi}$ is smaller than one, that is equivalent to assume $\lambda_t<\gamma+1$ as shown in \cite{corsi2013when}.}. It is also $\lambda\geq 1$ and $0\leq m\leq M$, by construction. 

Summarizing the main steps of the dynamics: (i) portfolio decisions at time $t-1$ determine the value of leverage depending on risk expectations, see Eqs. \ref{slow_d}, and (ii) affect the price dynamics between $t-1$ and $t$ because of the autoregressive coefficient $\bm{\Phi}(\lambda_{t-1},\Sigma^\omega_{d,t-1},\Sigma_{u,t-1}^\omega)$, of the return process, see Eqs. \ref{fast_d}; (iii) at time $t$, banks estimate asset risks by using past observations of prices, $\widehat{\Sigma}_{d,t}(\{\bm{r}_{t-1+k/n}\}_{k=1,2,...,n})$ and $\widehat{\Sigma}_{u,t}(\{\bm{r}_{t-1+k/n}\}_{k=1,2,...,n})$. Then, (iv) banks form new risk expectations $\Sigma^\omega_{d,t}$ and $\Sigma^\omega_{u,t}$, and (v) make new portfolio decisions at time $t$.

\paragraph{{\bfseries One dimensional setting}} In the following, we study also a reduced version of the model obtained by considering one bank and one risky investment. In this setting, the price dynamics is governed by the autoregressive process in Eq. \ref{return} with $N=M=1$. We refer to the investment return at time $s$ as $r_s$ and to its variance estimated by using observations $\{r_{t-1+k/n}\}_{k=1,2,...,n}$ as $\hat{\sigma}^2_t$. In this setting we lose the aspect related to the diversification and the portfolio problem reduces to find the optimal value of the leverage subject to the Value at Risk constraint, \ie
$$
\max_{\lambda_t}\lambda_t(\mu-r_L)\:\:\:\mbox{s.t.}\:\:\:\alpha\sigma_t^\omega\lambda_t\leq 1
$$
where $\sigma_t^\omega$ is the expected volatility of the investment. Then, the equation governing the portfolio decisions is $\lambda_t=(\alpha\sigma_t^\omega)^{-1}$ with $(\sigma^\omega_{t})^2 = \omega (\sigma^\omega_{t-1})^2+(1-\omega)\hat{\sigma}_t^2
$. We can reduce the slow component of the model dynamics to a single equation representing a one-dimensional map for the dynamic variable $\lambda_t$. The reduced model is specified as follows, 
\begin{equation}\label{n1}
\begin{cases}
\lambda_t=\left(\omega\frac{1}{\lambda_{t-1}^2}+(1-\omega)\:\alpha^2\:\widehat{Var}\left[\sum_{k=1}^n r_{t-1+k/n}\right]\right)^{-\frac{1}{2}}\\
r_s = \epsilon_s+\frac{\lambda_{t-1}-1}{\gamma}r_{s-1/n}\:\:\:~~~~~s=t-1+k/n,\:\:k=1,2,...,n
\end{cases}
\end{equation}
where 
$$
\widehat{Var}\left[\sum_{k=1}^n r_{t-1+k/n}\right]=\left(1+2\frac{\hat{\phi}_{t-1}(1-\hat{\phi}_{t-1}^n)}{1-\hat{\phi}_{t-1}}-2\frac{(n\hat{\phi}_{t-1}-n-1)\hat{\phi}_{t-1}^{n+1}+\hat{\phi}_{t-1}}{n(1-\hat{\phi}_{t-1})^2}\right)\frac{n\hat{\sigma}_\epsilon^2}{1-\hat{\phi}_{t-1}^2}
$$
is the maximum likelihood estimator of the variance of the return aggregated at the time scale of the portfolio decisions for generic $n>1$, with
$$
\hat{\phi}_{t-1}=\frac{\sum_{k=1}^n\:r_{t-1+k/n}r_{t-1+(k-1)/n}}{\sum_{k=1}^n\:r^2_{t-1+(k-1)/n}}
$$
$$
\hat{\sigma}_\epsilon^2=\frac{\sum_{k=1}^n(r_{t-1+k/n}-\hat{\phi}_{t-1}\:r_{t-1+(k-1)/n})^2}{n}
$$
the maximum likelihood estimators of the autoregressive coefficient $\phi_{t-1}\equiv\frac{\lambda_{t-1}-1}{\gamma}$ and the variance of the idiosyncratic noise $\sigma_\epsilon^2$ of the AR(1) process.

In Eq. \ref{n1} covariance stationarity for the AR(1) process is equivalent to the condition $\lambda\in[1,\gamma+1)$.

Since the slow variable evolves in time depending on averages over the fast variables, namely the variance of the random autoregressive process AR(1), this is a slow-fast random dynamical system.

In this setting, we can also consider the case $n=1$, i.e. time scale of portfolio decisions coincides with the one of portfolio rebalancing. In Appendix \ref{model1D-dt1}, we show that this case is similar to the one presented in \cite{aymanns2015dynamics} and, with a further assumption for the asset price dynamics, the two models coincide.

\bigskip

In the following we will consider three cases. 
\begin{itemize}
\item {\bf Exogenous risk expectations.} In the first case, we consider exogenous risk expectations, as studied in the paper \cite{corsi2013when}, thus we shall only summarize the main results obtained there.
\item {\bf Asymptotic deterministic limit of the slow-fast random dynamics.} The second case is the limit $n\rightarrow\infty$, \ie portfolio are rebalanced very actively between $t-1$ and $t$. In this limit the model becomes analytically tractable.
\item {\bf Two generic time scales.} The last is the case of finite $n$ and, for simplicity, we consider the reduced version of the slow-fast random dynamical system. We will present some numerical results.
\end{itemize}

\section{Exogenous risk expectations}\label{corsiresult}

We first consider the portfolio decision as completely exogenous, in the sense that expected diversifiable and systematic risks are exogenous parameters not depending on past asset price dynamics. This means removing the process of expectation formation in Eq. \ref{slow_d} and, as a consequence, considering only the fast component of the dynamics with no update of risk estimation. Since this case has already studied in \cite{corsi2013when}, we will not consider it in detail here but we summarize the findings. 

The authors show how changes in the constraints of the bank portfolio optimization (such as changes in the cost of diversification or changes in the micro-prudential policies) endogenously drive the dynamics of bank balance sheets, asset prices, and systemic risk. Specifically, a reduction of diversification costs, by increasing the level of diversification and hence relaxing the VaR constraint, allows the financial institutions to increase the optimal leverage and it also increases the degree of overlap, and thereby correlation, between the portfolios of financial institutions. This increases the feedback effects in the return dynamics that, in turn, increase the risk of investments and may induce steep growths (bubbles) and plunges (busts) of market prices. The authors provide a full analytical quantification of the multivariate feedback effects between investment prices and balance sheet dynamics induced by portfolio rebalancing in presence of asset illiquidity and show how the results crucially depend on the largest eigenvalue of the process in Eq. \ref{fast_d}. From a mathematical point of view, a transition from stationarity to non stationarity of the price dynamics occurs when the largest eigenvalue of $\bm{\Phi}$ becomes larger than one and it describes the dynamical instabilities of the financial system with exogenous risk expectations. Furthermore, the authors show that the bank size heterogeneity does not affect crucially the analytical results obtained in the homogenous case. On the contrary, bank size heterogeneity makes the financial system more unstable as compared to the homogeneous case because the maximum eigenvalue is larger for the heterogenous than for the homogeneous case, making the system closer to the transition between the stationary and the non-stationary dynamics.

\section{Asymptotic deterministic limit of the slow-fast random dynamics}\label{ds}

We now focus on the limit $n\rightarrow\infty$ which allows to approach the problem in a fully analytical way.  Since $n$ represents the number of times financial institutions rebalance their portfolios, it is a measure of how closely they are marked-to-market in the capital structure. The asymptotic limit $n\rightarrow\infty$ is equivalent to consider all financial institutions continuously marked-to-market in their capital structures. For a VaR constrained financial institution this means that the asset size is equal to the desired position with respect to the market value of the investments in such a way that the financial leverage is equal to its target during the portfolio holding period. Very active management of portfolio means frequent trading in financial market and thus can be accomplished only in absence of or with very low transaction costs and other trading frictions. For this reason, we can interpret the limit $n\rightarrow\infty$ as an ideal market without trading frictions.

In this limit, the estimator of the covariance matrix, $\widehat{\bm{\Sigma}}_t$, converges in probability to the covariance matrix $\widetilde{\bm{\Sigma}}$ for the returns $R_{i,t}$,
\begin{equation}\label{covendo}
\widehat{\bm{\Sigma}}_t\xrightarrow{\P}\widetilde{\bm{\Sigma}}= \widetilde{\Sigma}_d(\lambda_{t-1},\Sigma^\omega_{d,t-1},\Sigma^\omega_{u,t-1},\Sigma_\epsilon,\Sigma_f)\mathbb{I}+ \widetilde{\Sigma}_u (\lambda_{t-1},\Sigma^\omega_{d,t-1},\Sigma^\omega_{u,t-1},\Sigma_\epsilon,\Sigma_f)\bm{1}
\end{equation}
where $\Sigma_\epsilon$ and $\Sigma_f$ are the variance of the idiosyncratic noise $\epsilon_{i,s}$ and of the factor $f_s$, respectively, aggregated at the time scale of portfolio decisions\footnote{In the limit $n\rightarrow\infty$, the time scale of the portfolio rebalancing, \ie $\frac{1}{n}$, goes to zero as well as $\sigma_\epsilon^2$ and $\sigma_f^2$, but $\Sigma_\epsilon=\lim_{n\rightarrow\infty}n\:\sigma_\epsilon^2$ and $\Sigma_f=\lim_{n\rightarrow\infty}n\:\sigma_f^2$ remain finite.}, \ie $\Sigma_\epsilon=\lim_{n\rightarrow\infty}n\sigma_\epsilon^2$, $\Sigma_f=\lim_{n\rightarrow\infty}n\sigma_f^2$. 
$\widetilde{\Sigma}_d$ and $\widetilde{\Sigma}_u$ can be computed analytically, exploiting a result in \cite{corsi2013when}, and their expression is in Eq. \ref{SigmaAsymptotic} of Appendix \ref{covarianceAppendix}.

Substituting the estimator with the covariance matrix of Eq. \ref{covendo}, the portfolio dynamics is described by a three-dimensional deterministic map, which in implicit form is:
\begin{equation}\label{dyn_s}
\begin{cases}
\left(\Sigma^\omega_{d,t-1}\right)^{\frac{1}{3}}\left(\alpha^{\frac{2}{3}}\left(\frac{\mu-r_L}{2c}\right)^{\frac{1}{3}}\lambda_t\right)^{-1}-\left(\frac{1}{\alpha^2\lambda_t^2}-\Sigma^\omega_{u,t-1}\right)^{\frac{2}{3}}=0 \\
\Sigma^\omega_{d,t}=\omega\Sigma^\omega_{d,t-1}+(1-\omega)\widetilde{\Sigma}_d(\lambda_{t-1},\Sigma^\omega_{d,t-1},\Sigma^\omega_{u,t-1},\Sigma_\epsilon,\Sigma_f)\\
\Sigma^\omega_{u,t}=\omega\Sigma^\omega_{u,t-1}+(1-\omega)\widetilde{\Sigma}_u(\lambda_{t-1},\Sigma^\omega_{d,t-1},\Sigma^\omega_{u,t-1},\Sigma_\epsilon,\Sigma_f).
\end{cases}
\end{equation}
The map of Eq. \ref{dyn_s} is the so-called \emph{deterministic skeleton} of the financial model obtained by removing the sources of stochasticity.

\subsection{Deterministic skeleton of the financial system}\label{dynamicalsystem}
\begin{table}[t]
\begin{center}
\begin{tabular}{|p{2cm}|p{5cm}|p{1.5cm}|}
\hline
{\bfseries Notation} & {\bfseries Description} & {\bfseries Value} \\
\hline
$M$ & total number of investments & $60$ \\
\hline
$N$ & number of banks & $30$ \\
\hline
$\mu-r_L$ & Net Interest Margin of a bank & $0.08$ \\
\hline
$\gamma$ & asset liquidity & $100$ \\
\hline
$\sqrt{\Sigma_\epsilon}$ & exogenous idiosyncratic volatility at the time scale of portfolio decisions & $0.03$ \\
\hline
$\sqrt{\frac{\Sigma_f}{\Sigma_\epsilon}}$ & ratio between exogenous volatility of systematic factor at the time scale of portfolio decisions, \ie $\Sigma_f$, and $\Sigma_\epsilon$ & $0.1$\\
\hline
$A_0$ & Initial asset size for each bank & $100$ \\
\hline
$E_0$ & Initial equity for each bank & $\frac{A_0}{\lambda_0}$ \\
\hline
$c$ & cost of diversification w.r.t. the initial equity $E_0$ & $0.1$ \\
\hline
$\alpha$ & quantile of $P_{VaR}$ & $1.64$ \\
\hline
$\lambda_0,\Sigma^\omega_{d,0},\Sigma^\omega_{u,0}$ & chosen randomly in the domain &  \\
\hline
\end{tabular}
\end{center}
\caption{Simulation parameters for the model. The most important parameters in our analysis are $\omega$, $c$ and $\alpha$. We will specify $\omega$ below while $\alpha$ and $c$ are set as in the Table if it is not specified differently in the subsequent figures. Since we study the long-run dynamics, the initial conditions are not relevant.}
\label{table_parameters}
\end{table}

We analyze the model in the asymptotic limit via a dynamical systems approach by investigating the map in Eq. \ref{dyn_s}. Table \ref{table_parameters} shows the values of the parameters used to illustrate the main properties of the model.
\begin{figure}[t]
{\includegraphics[width=0.88\textwidth]{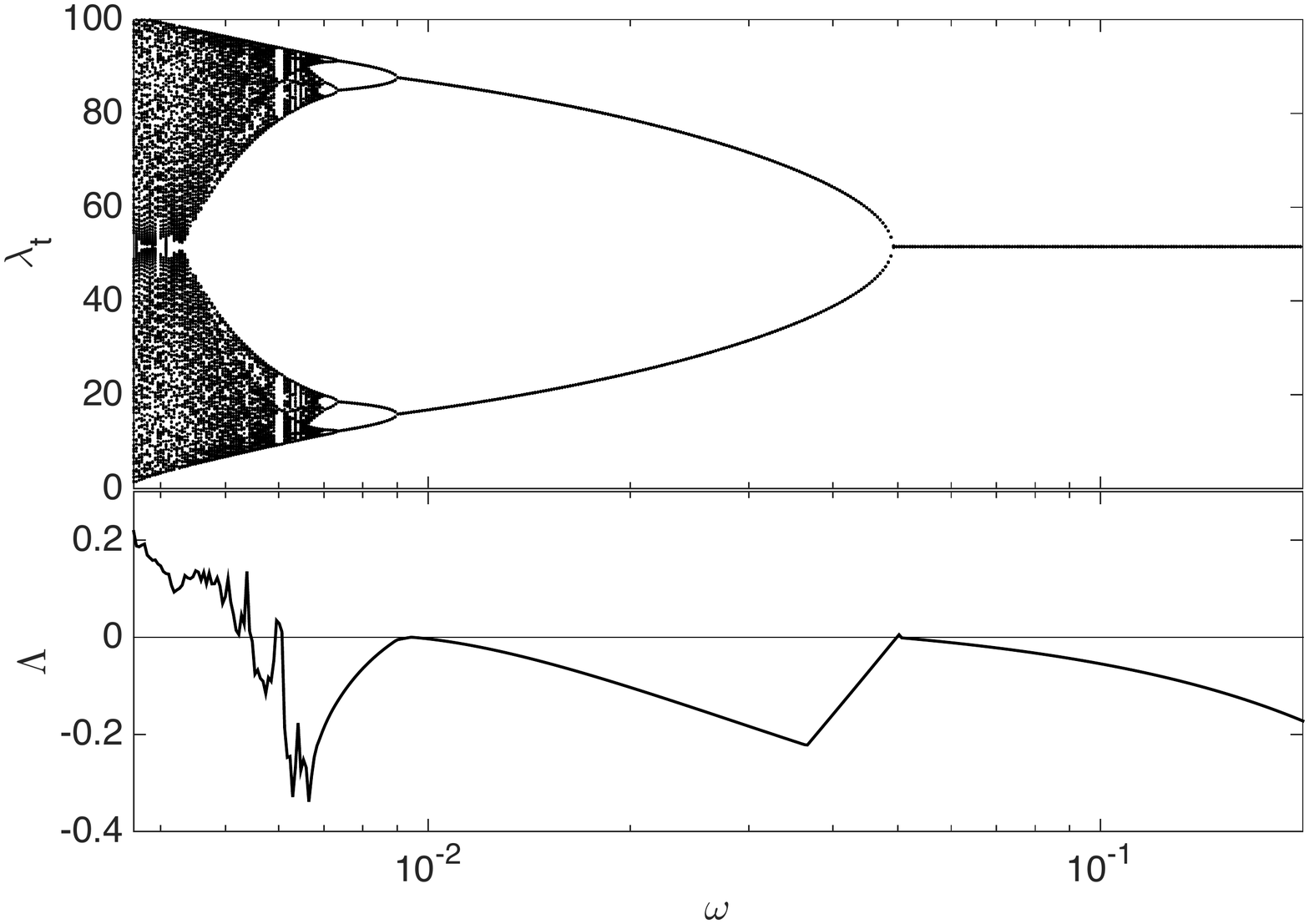}}
\caption{\footnotesize{Top panel: bifurcation diagram associated with $\lambda_t$ for the analytical mapping of Eq. \ref{dyn_s} as a function of the memory $\omega$. Bottom panel: the largest Lyapunov exponent $\Lambda$ associated with the bifurcation diagram. $\Lambda$ is obtained numerically via the method of the orbits \cite{wolf1985determining}.}}
\label{bifomega}
\end{figure}

A comprehensive insight of the skeleton dynamics is obtained by looking at the bifurcation diagram in the top panel of Figure \ref{bifomega}. The bifurcation diagram of a map shows the long run dynamics after a transient time has passed as a function of a parameter. For our model we can use as the bifurcation parameter either the memory parameter $\omega$ or the VaR parameter $\alpha$, or the diversification cost $c$. Later, we will show the model dynamics depending on $\alpha$ and $c$ to have more insights regarding policy implications.

The top panel of figure \ref{bifomega} shows the bifurcation diagram as a function of $\omega$. Decreasing the memory parameter, a period-doubling cascade to chaos occurs. In other words the deterministic skeleton of the financial system shows cycles of increasing complexity with more and more periods until chaos occurs\footnote{In this analytical approach to the study of dynamics of the financial system, the domain for leverage and risk expectations is $\{\lambda,\Sigma^\omega_{d},\Sigma_u^\omega\}\in[1,\gamma+1)\times\mathbb{R}^+\times\mathbb{R}^+$. In particular, bounds for leverage are equivalent to the conditions of stationarity for the process of Eq. \ref{fast_d}.}. The signature of the chaotic behaviour is the positive Lyapunov exponent associated with it. The bottom panel of Figure \ref{bifomega} shows the estimated Lyapunov exponent $\Lambda$ as a function of $\omega$ and positive values of $\Lambda$, observed  for small $\omega$, signal the presence of  deterministic chaos.

Note that the unit of $\omega$ in Figure \ref{bifomega} is the inverse of the time scale of the portfolio decisions, see Eq. \ref{expectations2}. By using as unit the time scale of portfolio rebalancing, the time scale becomes $\tilde{\omega}=\omega^\frac{1}{n}$. For example, if banks rebalance their portfolios at daily frequency but the portfolio decisions are taken once a month, a memory parameter $\tilde{\omega}=0.97$ at the daily time scale corresponds to an effective memory of $\tau\approx33$ days. The corresponding value of the memory parameter at the monthly time scale is $\omega\approx0.4$. Thus small values of $\omega$, which in the bifurcation diagram corresponds to cycles or chaos, are associated with values of $\tilde \omega$ close to $1$.

One can understand the behaviour in Figure \ref{bifomega} by considering that, when memory is large enough, banks learn from past history and, at least in the long-run, they make exact forecasts of future risks. Thus large memory stabilizes the financial system and portfolio decisions about leverage and diversification reach a fixed point equilibrium. When $\omega$ decreases below a given threshold, risk expectations with smaller memory break the fixed point equilibrium by inducing cycles of period two. Indeed, a shorter memory has the potential of creating panic-induced fall in portfolio holdings reflecting sudden decrease in leverage and diversification. When the adopted financial leverage and the overlap between banks' portfolios are high, the price impact of the portfolio rebalancing increases significantly the observed risks. Small memory in risk expectations tends to overestimate future risks and as a consequence banks reduce suddenly their positions. Hence, leverage and diversification suddenly decrease. The opposite situation occurs when leverage and diversification are small due to previously overestimated risk. The 2-period cycle is characterized by a mismatch between the expectation and the realization of portfolio volatility and leverage cycles reflect the mismatch between the banks' perceived risk and the `true' risk. 

When the memory decreases further, our model predicts cycles of larger periods and, eventually, a chaotic dynamics. Notice that the amplitude of both periodic and chaotic leverage cycles increases by decreasing $\omega$. When $\omega>\omega^*$, $\lambda$ oscillate within $[1,\gamma+1)$ and the covariance stationarity of the return is guaranteed, while when $\omega<\omega^*$ the process becomes non-stationary. Hence, $\omega^*$ corresponds to a transition from stationarity to non-stationarity for the model dynamics. In the left panel of Figure \ref{cha}, we show the contour map of $\omega^*$ as a function of the square root of the idiosyncratic variance $\sqrt{\Sigma_\epsilon}$ and the total number of investments $M$. The larger is $\sqrt{\Sigma_\epsilon}$, the smaller is the value of $\omega^*$: this means that larger exogenous risk reduces the value of the financial leverage because of the Value at Risk constraint and, as a consequence, the feedback effects of leverage targeting. More surprisingly, the value of $\omega^*$ increases with $M$, that is introducing financial innovations and new instruments requires investors with larger memory in forming risk expectations to keep the system in stationary equilibrium. Indeed, a larger number $M$ of available asset investments leads to an underestimation of risk because of larger portfolio diversification which is followed by an increase of the financial leverage. Then, the combined effects of the overlap of banks' portfolios with the price impact of the leverage targeting make the financial system more unstable.

\subsection{Bifurcation analysis}
\begin{figure}[t]
\centering
{
	\includegraphics[width=0.51\textwidth]{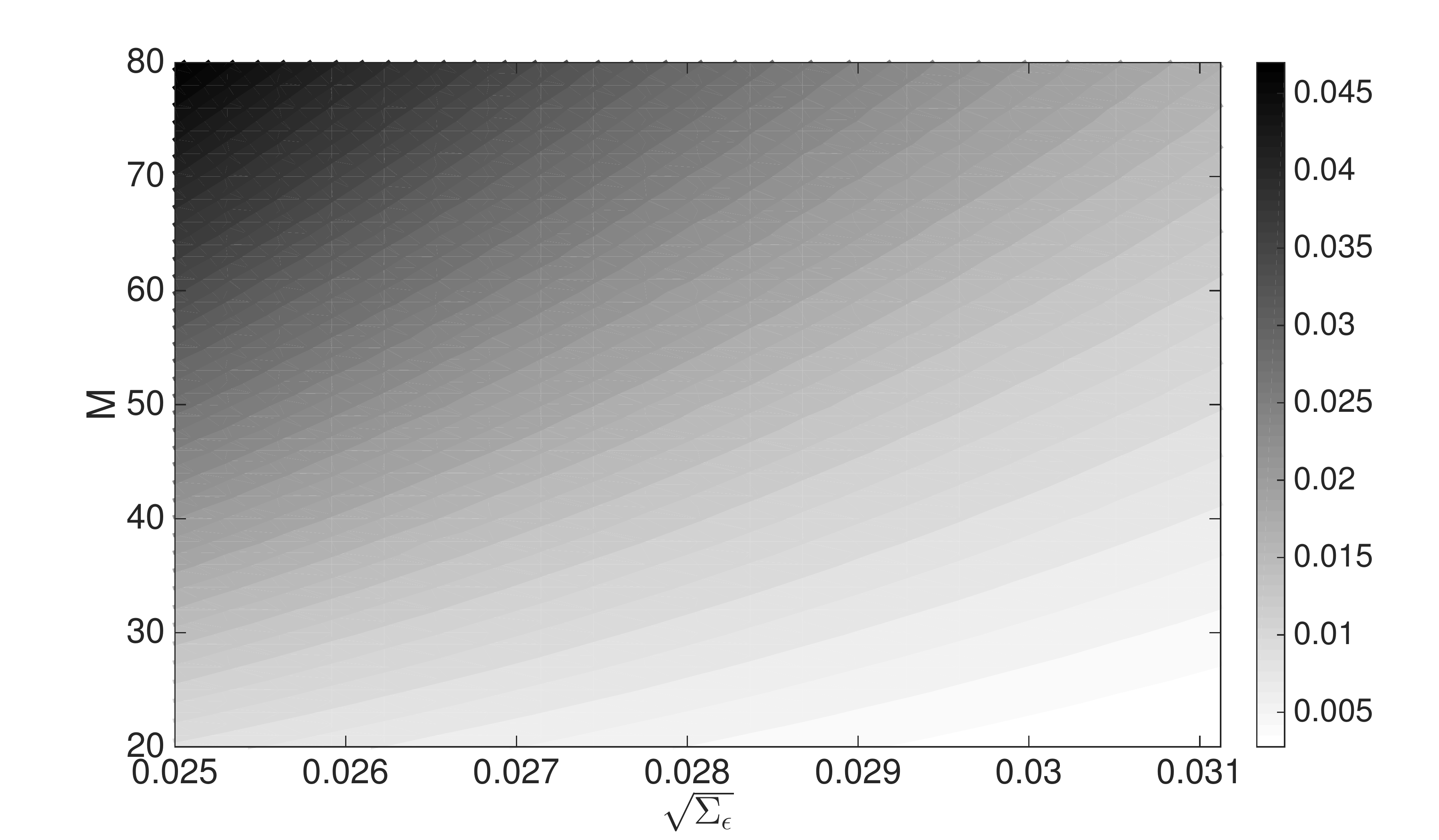}}
{
	\includegraphics[width=0.47\textwidth]{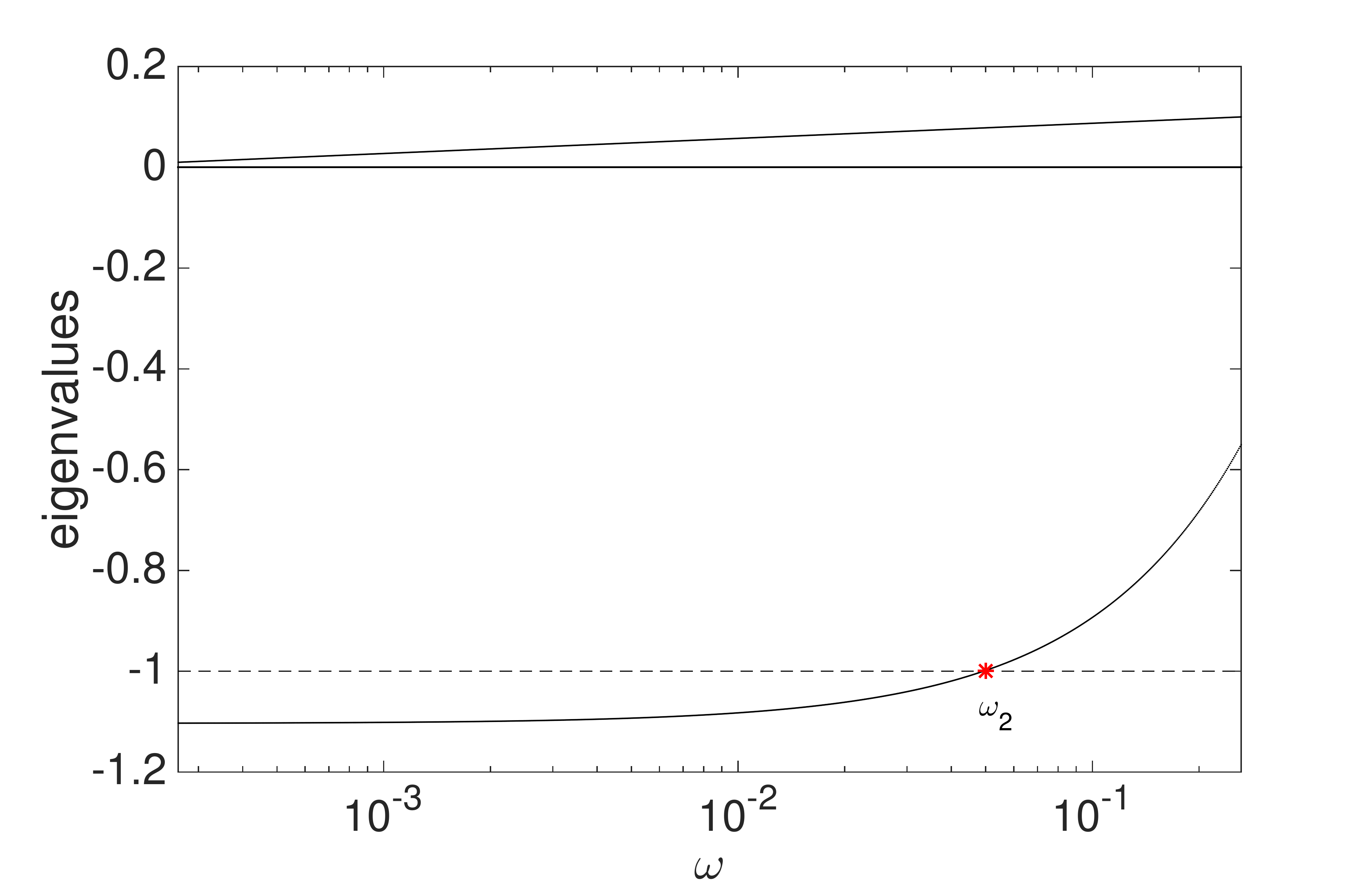}}
\caption{\footnotesize{Left panel: contour map of $\omega^*$ (\ie the value of $\omega$ at which the transition to non stationarity occurs) as a function of $\sqrt{\Sigma_\epsilon}$ and the number of investments $M$. Here $\Sigma_f=0$ and the other parameters are specified as in Table \ref{table_parameters}. Right panel: eigenvalues of the Jacobian for the map of Eq. \ref{dyn_s} as a function of $\omega$ and parameters specified as in Table \ref{table_parameters}.}}
\label{cha}
\end{figure}
The breaking of fixed point equilibrium occurs by a period-doubling bifurcation or flip bifurcation (see for example \cite{crawford1991introduction}). In the case of $\omega$ as bifurcation parameter, we refer to the value of the memory parameter for which the period-doubling bifurcation occurs as $\omega_2$. The type of bifurcation can be analyzed by studying the Jacobian associated with the map of Eq. \ref{dyn_s}. The map describes a three-dimensional dynamical system but one eigenvalue of the Jacobian is identically null which suggests that the two equations associated with the process of expectation formation in Eq. \ref{dyn_s} are not independent at the first order in the expansion. The positive eigenvalue describes the eigenspace associated with the independent component of these equations. The dynamics in this eigenspace is characterized by a fixed point since the corresponding eigenvalue is inside the unit circle for all $\omega$. The third eigenvalue is instead negative. The right panel of Figure \ref{cha} shows the non vanishing eigenvalues of the Jacobian as a function of $\omega$. When the negative eigenvalue hits the critical value $-1$ for $\omega=\omega_2$, a period-doubling bifurcation occurs. After that, the attractor for the dynamics is not the fixed point anymore and a new solution appears describing the 2-period cycles. The new solution is the fixed point for the map of Eq. \ref{dyn_s} iterated twice. In a self-similar way, the period-doubling bifurcation repeats again but for the iterated map giving rise to the period-doubling cascade to chaos. The phenomenon is very general, was noticed for the first time by studying the logistic map \cite{may1976simple} and was described in \cite{feigenbaum1978quantitative}.

\begin{figure}[t]
\centering
{
	\includegraphics[width=0.48\textwidth]{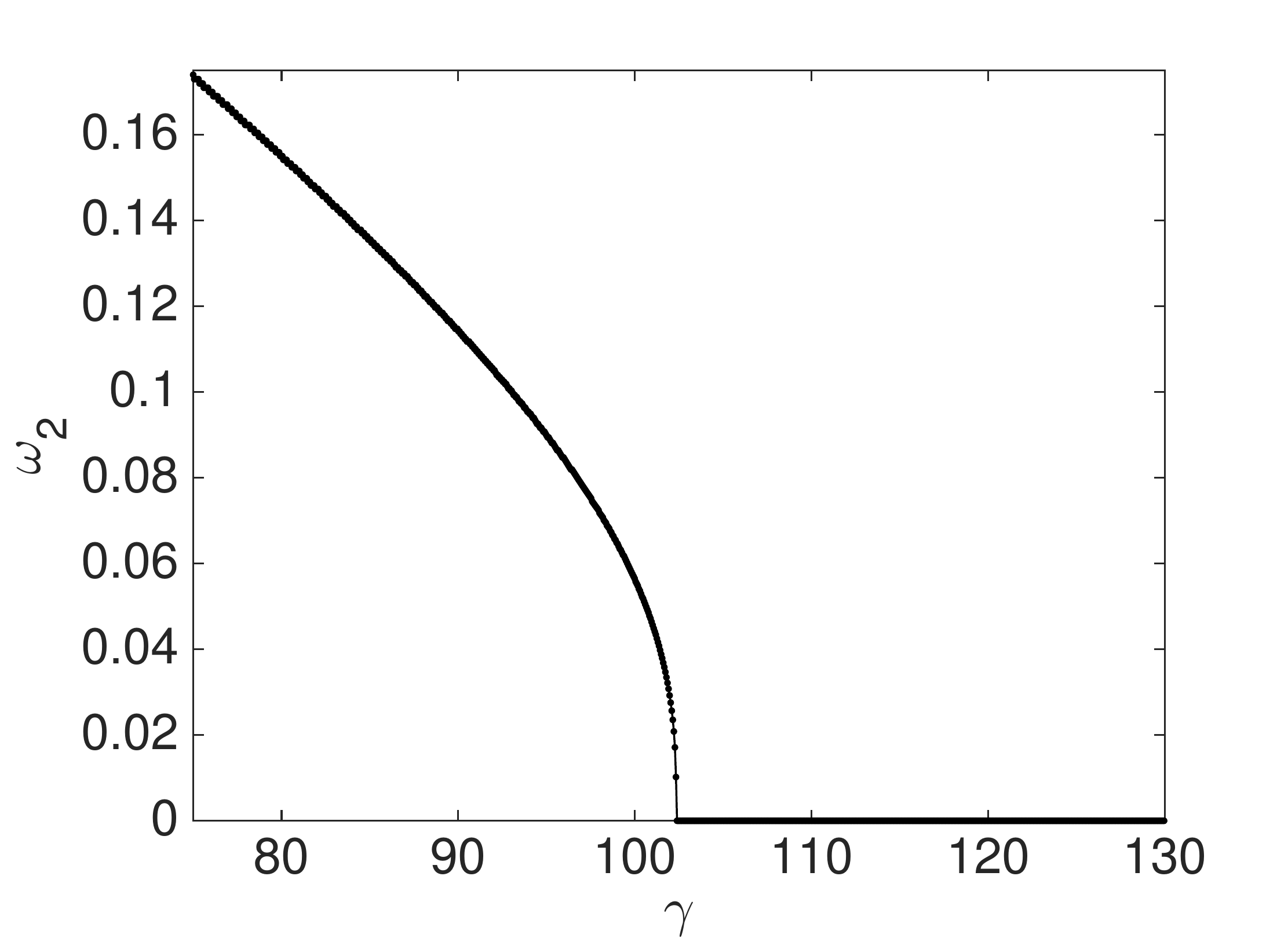}}
{
	\includegraphics[width=0.45\textwidth]{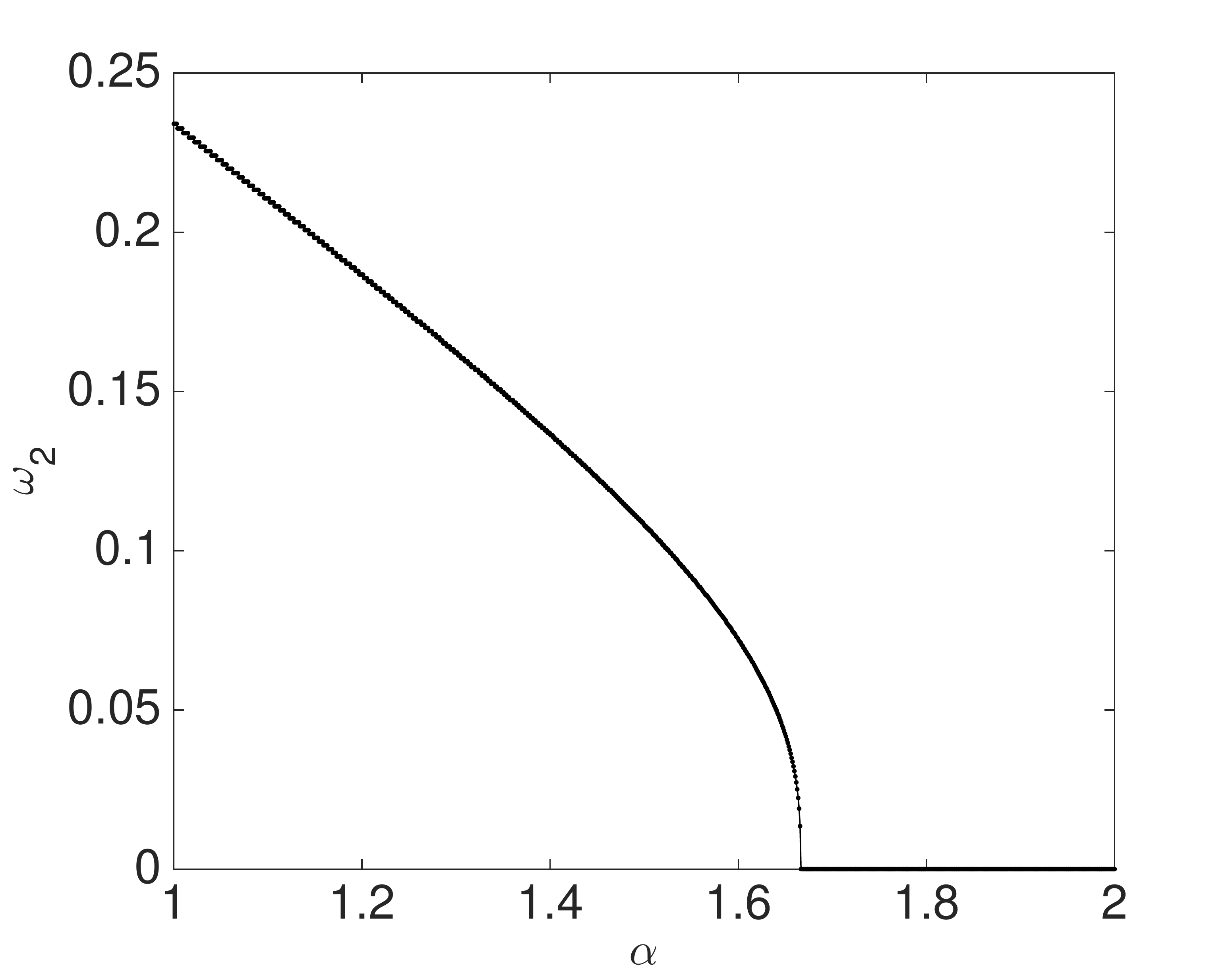}}
{
	\includegraphics[width=0.8\textwidth]{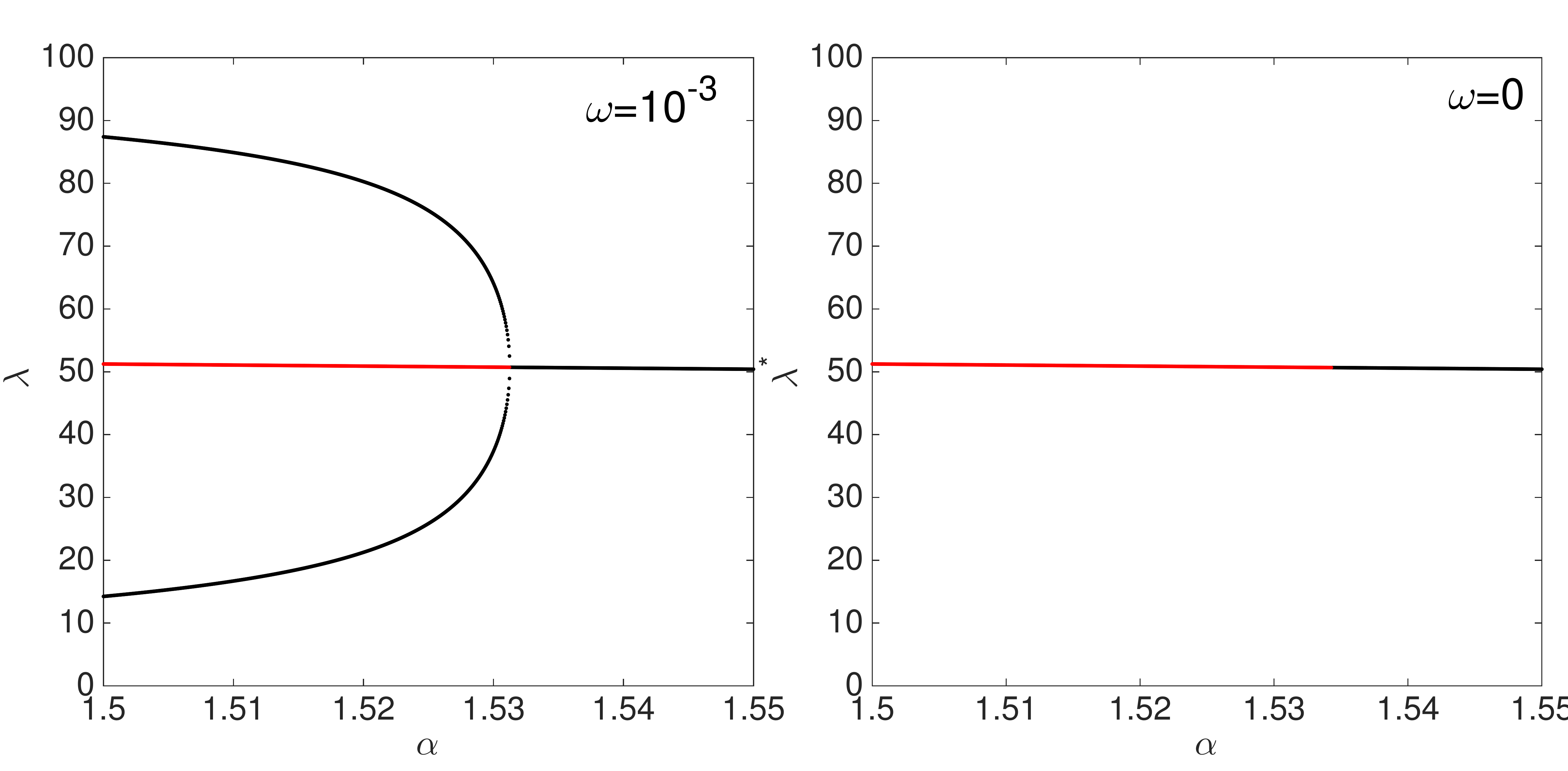}}
\caption{\footnotesize{ Top panels:: the bifurcation parameter $\omega_2$ corresponding to the first period-doubling bifurcation as a function of market liquidity of assets $\gamma$ (left) and quantile $\alpha$ of the $P_{VaR}$ (right). Bottom panels: fixed point analysis of the analytical mapping in Eq. \ref{dyn_s}. Black dots represent attractive stable points (fixed points or stable cycles of period 2) while red dots represent unstable fixed points. Bottom left panel: adaptive expectations with non zero memory ($\omega=10^{-3}$). Bottom right panel: \emph{naive} expectations characterized by zero memory ($\omega=0$).}}
\label{c2}
\end{figure}

We also study the stability properties of the financial system by analyzing the value of $\omega_2$ as a function of the other model parameters. In the top left panel of Figure \ref{c2}, we show the relation between the value $\omega_2$ and the liquidity parameter $\gamma$. The larger is the market liquidity of risky investments, the less important are the effects related to the expectation feedbacks. Hence, the larger is the market liquidity, the more stable is the dynamics of the financial system. Intuitively, when the price impact of the portfolio rebalancing to target the financial leverage does not affect significantly the investment prices, then volatility and portfolios do not fluctuate much. Liquid markets are more stable because endogenous impact of market players' strategy affects less the risk \cite{danielsson2012endogenous,danielsson2012endogenous_extreme}. Finally notice that, since the transition values, such as $\omega_2$, depend on liquidity, an abrupt change in liquidity can drive the market from a stable to an unstable regime. In the top right panel of Figure \ref{c2}, we show the relation between $\alpha$ and $\omega_2$. The decreasing behaviour indicates that when Value at Risk is less stringent, the financial market becomes dynamically stable if the memory used by agents to build expectations is longer. 

In the bottom panels of Figure \ref{c2}, we show the fixed point analysis of the analytical mapping in Eq. \ref{dyn_s} in the neighbourhood of the first bifurcation point. We compare the case of adaptive expectations of risk described in Eq. \ref{expectations} with naive expectations, \ie the case of zero memory ($\omega=0$ in Eq. \ref{expectations}). The bottom left panel shows the stable points of the map (black dots) for the variable $\lambda$ when memory is different from zero, \ie the fixed point of the map in Eq. \ref{dyn_s} or the periodic orbit that is the fixed point for the map iterated twice. When the bifurcation parameter $\alpha$ decreases below the threshold determining the period-doubling bifurcation, the new solution for the map iterated twice appears and the fixed point of the map becomes unstable (red dots). This behaviour holds for any $\omega$ different from zero. The bottom right panel of Figure \ref{c2} shows the case of naive expectations of risk. At the bifurcation point, no solutions for the map iterated twice exist. However, the fixed point is not attractive anymore. In this case, the dynamics of the system is cyclical but characterized by amplitude increasing in time\footnote{Clearly, the map of Eq. \ref{dyn_s} makes sense as long as the variables are in the domain.}. Hence, when financial agents behave naively in risk forecasting, the breaking of fixed point equilibrium corresponds to a transition from stationary to non stationary dynamics for the financial system.

\subsection{Policy implications}
\begin{figure}[t]
{
	\includegraphics[width=0.49\textwidth]{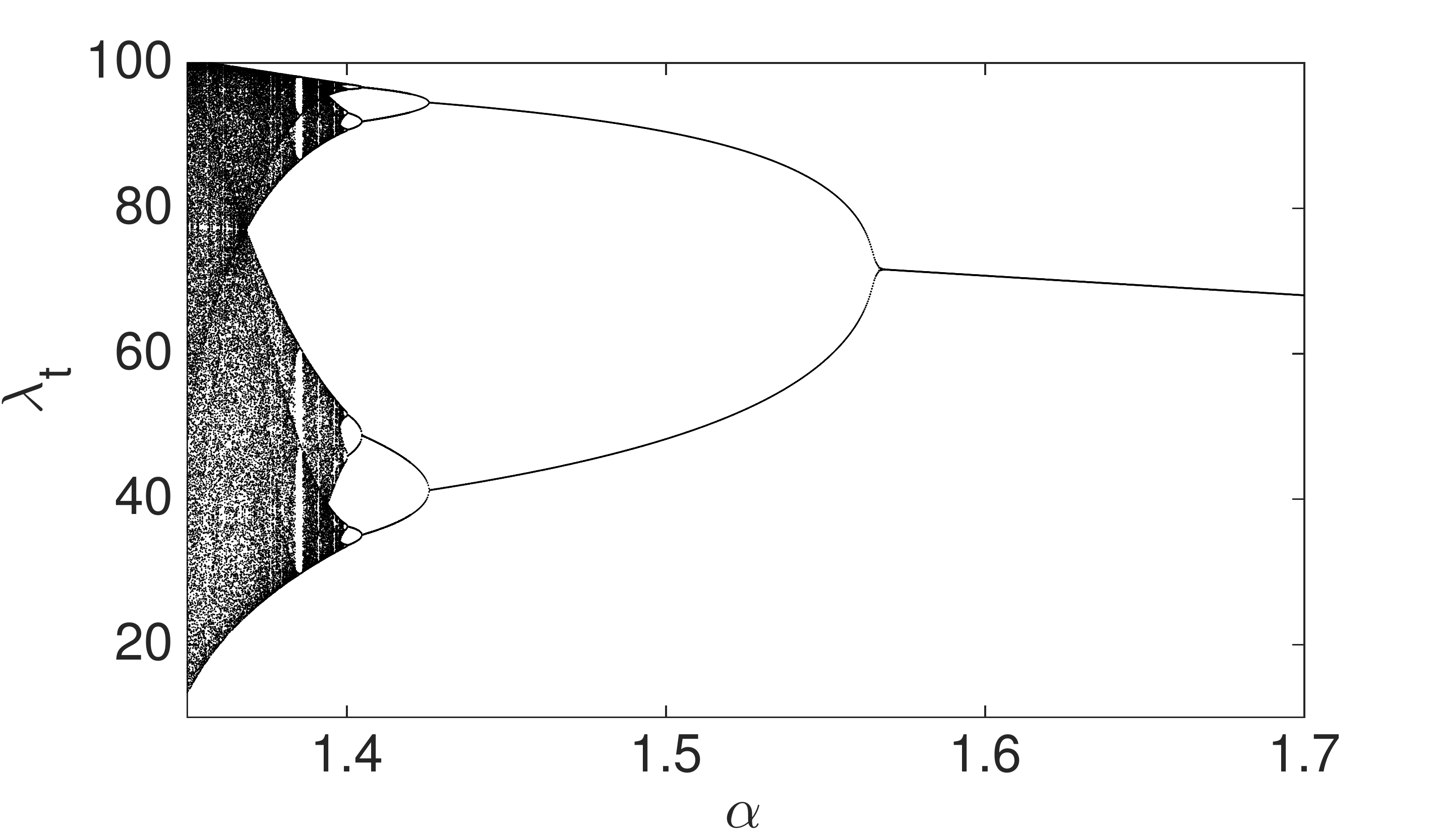}}
{
	\includegraphics[width=0.49\textwidth]{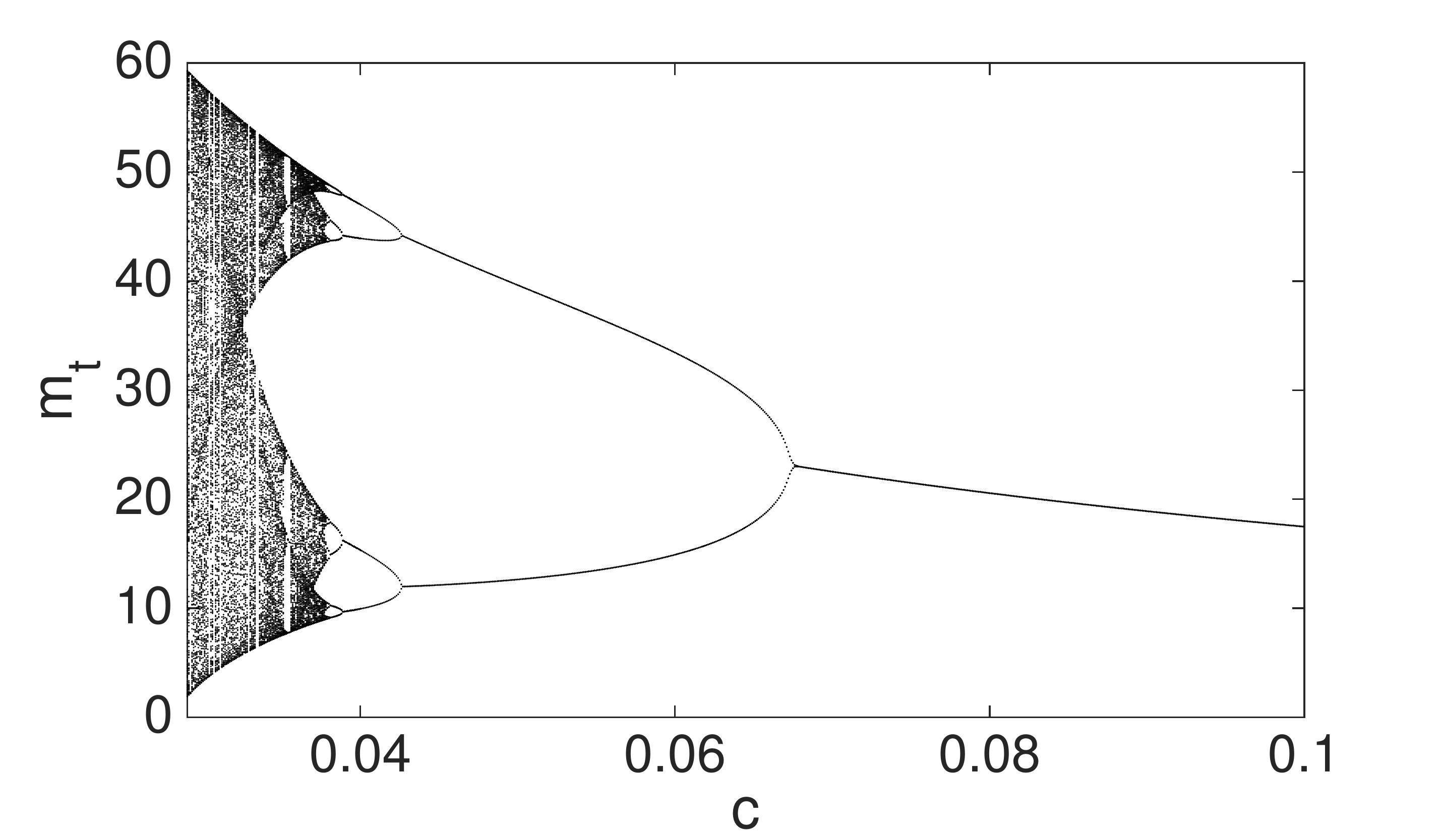}}
\caption{\footnotesize{Left panel: bifurcation diagram associated with $\lambda_t$ for the analytical mapping \ref{dyn_s} as a function of $\alpha$. $\omega$ is set equal to $0.1$. Right panel: bifurcation diagram for $m_t$ as a function of diversification cost $c$. $\omega$ is set equal to $0.1$.}}
\label{bif}
\end{figure}
From the point of view of financial policy, our results can be summarized as follows.
\begin{itemize}
\item Decreasing the parameter of the VaR, thus allowing a less stringent capital constraint, makes financial system more unstable. For example, in the described financial system characterized by an average investment volatility of $3\%$ in the holding period of the portfolio, diversification costs about $10\%$ of the equity value and a memory in risk expectations equal to $\omega=0.1$, relaxing the probability of loss $P_{VaR}$ from $5\%$ to $6\%$ ($\alpha=1.64\rightarrow1.5$) breaks the fixed point equilibrium inducing leverage cycles. Furthermore, allowing probability of loss associated with Value at Risk of about $7-8\%$ leads the financial system towards chaotic evolution, see the left panel of Figure \ref{bif}.
\item Decreasing the cost of diversification tends to increase the dynamical instability of the financial system. Let us notice in the right panel of Figure \ref{bif} that decreasing diversification cost, e.g. from $10\%$ to $5\%$ of equity value, breaks systemic equilibrium. A similar effect is obtained by introducing financial innovations and new instruments which increase the number of available asset investments $M$, inducing an underestimation of risk because of larger diversification, see Figure \ref{cha}.
\item Imposing to financial institutions operating in the market to adopt larger memory in their expectation schemes for risk forecasting has always stabilizing effects on market stability. Not surprisingly, increasing $\omega$ means to take into account more information about the system, converging asymptotically to the fixed point equilibrium. However, larger memory means slower reaction to risk news. What could seems meaningless from a micro-prudential point of view, as it is if adopted only by one or few financial institutions, becomes convenient in terms of dynamical stability when all institutions behave similarly. Thus, our analysis suggests that an appropriate financial policy is necessary to promote a systemic behaviour which prefers smooth adjustments driven by more information to overreactions driven by news.
\end{itemize}

\subsection{One dimensional analysis}
\begin{figure}[t]
\centering
{
	\includegraphics[width=0.49\textwidth]{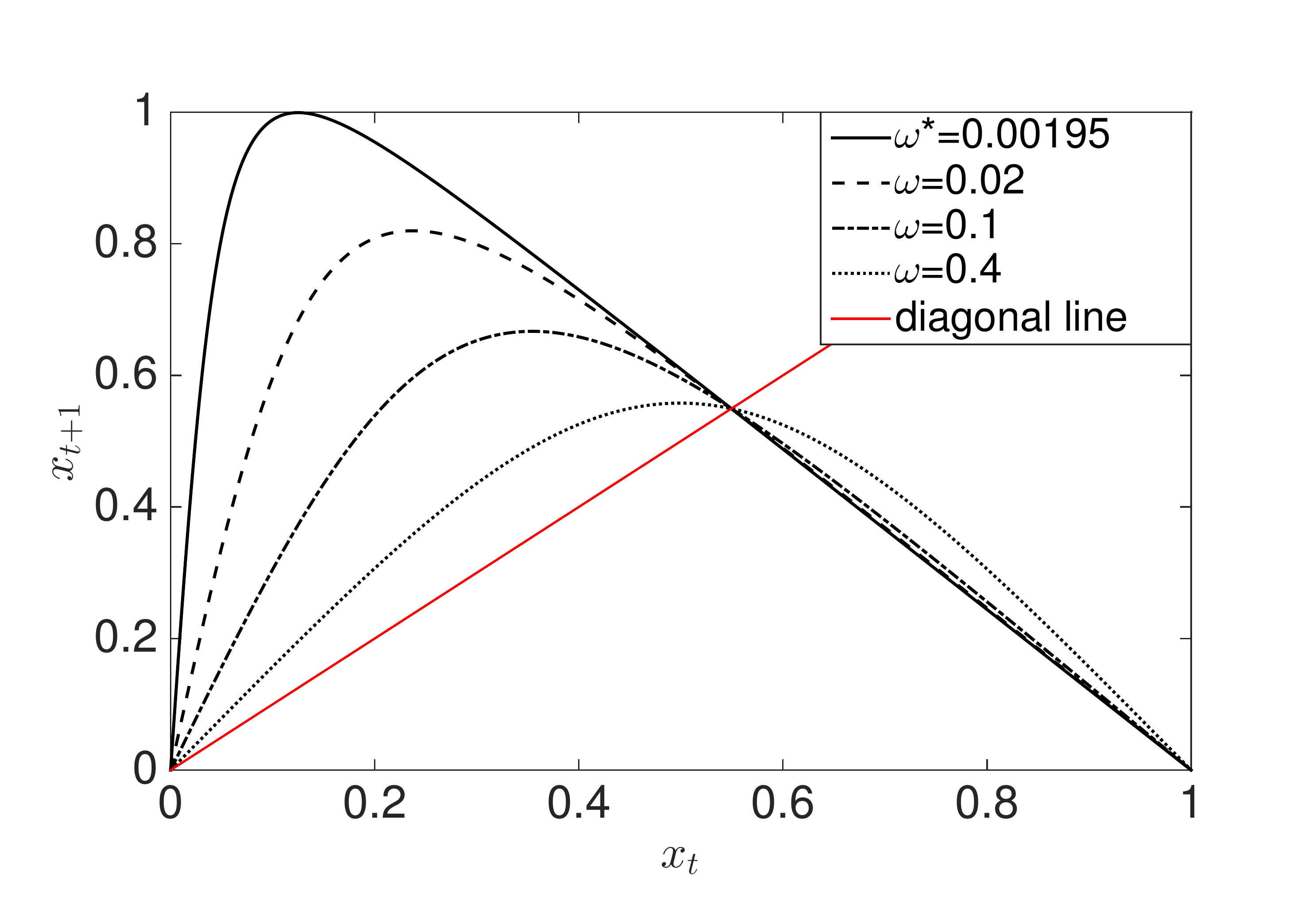}}
{
	\includegraphics[width=0.49\textwidth]{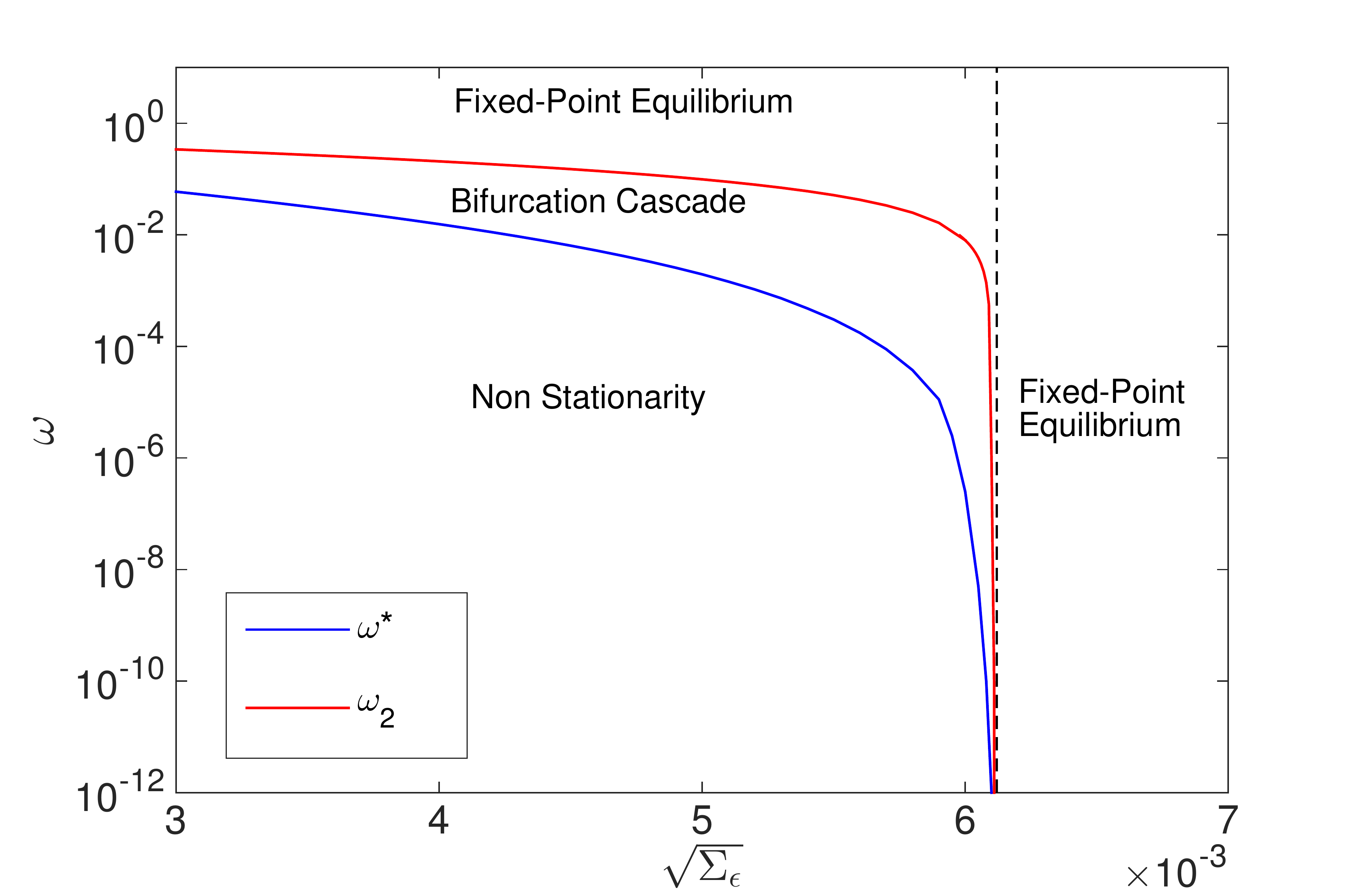}}
\caption{\footnotesize{Left panel: shape of the one dimensional map in Eq. \ref{MAPPAVERA} with domain rescaled to unitary interval, \ie we consider the linear transformation $x\equiv\frac{\lambda-1}{\gamma}$. The other parameters in Eq. \ref{MAPPAVERA} are: $\alpha=1.64$, $\gamma=100$, $\sqrt{\Sigma_\epsilon}=0.005$. Right panel: the value of the memory parameter at which the first period-doubling bifurcation occurs, \ie $\omega_2$ (red line), and the value at which the transition to non stationarity occurs, \ie $\omega^*$ (blue line), as a function of $\sqrt{\Sigma_\epsilon}$. The other parameters are $\alpha=1.64$ and $\gamma=100$.}}
\label{fig1D}
\end{figure}

The dynamics of  the reduced model in the limit $n\rightarrow\infty$ is governed by the following one dimensional map for $\lambda_t$,
\begin{equation}\label{MAPPAVERA}
\lambda_t=f(\lambda_{t-1}; \omega, \alpha, \gamma, \Sigma_\epsilon)=\left(\omega\frac{1}{\lambda_{t-1}^2}+(1-\omega)\:\alpha^2\left(1+2\frac{\lambda_{t-1}-1}{\gamma-\lambda_{t-1}+1}\right)\frac{\Sigma_\epsilon}{1-\left(\frac{\lambda_{t-1}-1}{\gamma}\right)^2}\right)^{-\frac{1}{2}}
\end{equation}
where $\lambda\in[1,\gamma+1)$ and $\Sigma_\epsilon=\lim_{n\rightarrow\infty}n\sigma_\epsilon^2$ is the variance of the idiosyncratic noise at the slow time scale of portfolio decisions, see Eq. \ref{n1}. In Appendix \ref{covarianceAppendix}, we show how to compute analytically the variance of the aggregated return $\sum_{q=1}^n r_{t-1+q/n}$ in the limit $n\rightarrow\infty$ to obtain Eq. \ref{MAPPAVERA}. The map $f$ in explicit form provides valuable analytical insights into the dynamical behavior of the system. The fixed point of the map, \ie $\lambda^*=f(\lambda^*; \omega, \alpha, \gamma, \Sigma_\epsilon)$, corresponds to the asymptotic equilibrium when $f'(\lambda^*; \omega, \alpha, \gamma, \Sigma_\epsilon)$ lies inside the unit circle. When $f'(\lambda^*; \omega, \alpha, \gamma, \Sigma_\epsilon)$  hits the value $-1$, the period-doubling bifurcation occurs. The left plot of Figure \ref{fig1D} shows the shape of the map, under the linear transformation $x\equiv\frac{\lambda-1}{\gamma}$, as a function of $\omega$. Notice that the value of the fixed point does not depend on $\omega$ and the maximum of the map is a decreasing function of $\omega$. Furthermore, the value $\omega^*$ at which the transition to non stationarity occurs corresponds to the value of $\omega$ at which the maximum of $f$ is equal to $\gamma+1$ (or the unit value when we rescale the invariant interval of the map as in Figure \ref{fig1D}). 

It is interesting to notice that the complex behavior of the period-doubling cascade to chaos may occur when the exogenous variance of the idiosyncratic noise is below a threshold value. In the right panel of Figure \ref{fig1D} we show $\omega_2$ and $\omega^*$ as a function of $\sqrt{\Sigma_\epsilon}$. When we approach the threshold (black dotted line), $\omega_2$ and $\omega^*$ tend to coincide and both values become arbitrarily low. Above the threshold, no breaking of the fixed point equilibrium occurs. The curves describing $\omega_2$, $\omega^*$, and the threshold partition the parameter space in regions characterized by specific dynamical properties, see Fig. \ref{fig1D}.

\section{Two generic time scales}\label{fluctuations}
The results of Section \ref{ds} are obtained in the asymptotic limit $n\rightarrow\infty$ corresponding to a situation where financial institutions are continuously marked-to-market in their capital structure. In reality, there are several types of frictions in trading operations which prevent continuous marking-to-market. It is interesting to investigate how the results obtained via the dynamical systems approach change when $n$ is finite. From a policy perspective this corresponds to investigate the role of frictions (e.g. transaction taxes) on the stability of the financial system.

\subsection{Perturbation analysis}\label{pa} 
For large but finite $n$, the estimator $\hat{\bm{\Sigma}}_t$ of the covariance matrix fluctuates around the  covariance matrix $\widetilde{\bm{\Sigma}}$. These fluctuations in risk estimation affect portfolio decisions about leverage and diversification and, as a consequence, also the dynamics of the financial system. Here, we want investigate how much fluctuations in risk estimations affect the deterministic skeleton of financial system. In the following, we present an analytical argument to answer this question. Then, in Subsection \ref{1D} we justify it by comparing the analytical results with Monte Carlo simulations of the reduced version of the model. 
\begin{figure}[t]
{\includegraphics[width=0.75\textwidth]{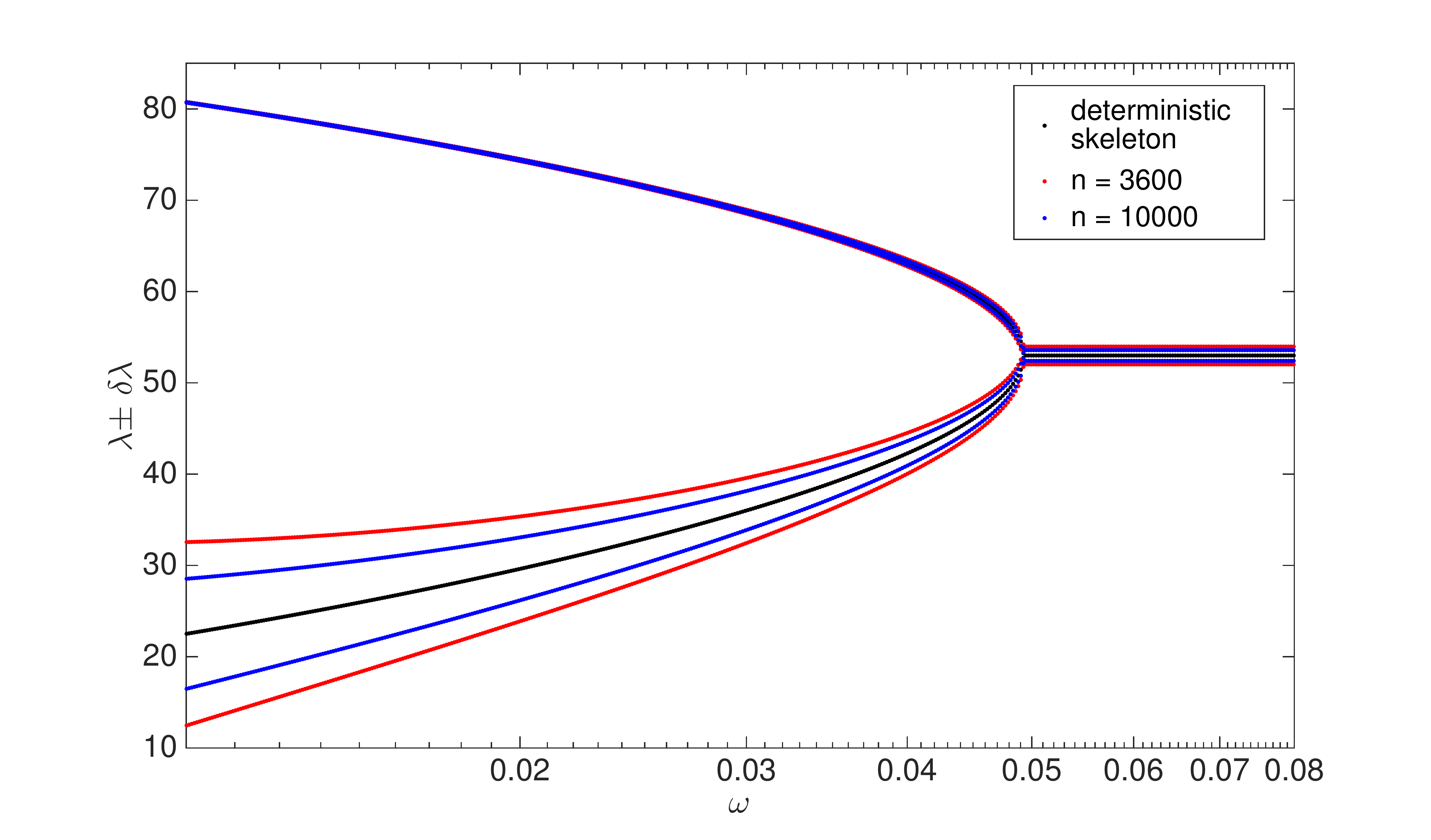}}
\caption{\footnotesize{Bifurcation diagram of the leverage as a function of $\omega$ and perturbations around the deterministic skeleton of the map \ref{dyn_s} for different values of $n$ obtained according to the perturbation analysis described in Subsection \ref{pa}.}} \label{fluctuations_fig}
\end{figure}

The asset return dynamics is governed by the VAR process of Eq. \ref{fast_d}. By estimating the autoregressive process between $t-1$ and $t$ when $n$ is finite, financial institutions can obtain the exogenous components of the VAR (1) process, \ie the residuals $\{\epsilon_{i,s},f_s\}_{s=t-1+k/n,\:k=1,2,...,n}^{i=1,...,M}$\footnote{From a mathematical point of view, the process \ref{fast_d} is equivalent to the process for the endogenous components of returns, see Equation \ref{endo2}. As shown in \cite{corsi2013when}, it can be decomposed in $M+1$ independent AR(1) process. The residuals of Eq. \ref{fast_d} are simply linear combinations of the residuals associated with the $M+1$ AR(1) processes.}. The estimator of the idiosyncratic and systematic risks at the time scale of portfolio rebalancing is the realized variance of the residuals,
\begin{equation}\label{realized_v}
\hat{\sigma}_{\epsilon}^2=\frac{1}{n-1}\sum_{k=1}^{n}\epsilon_{i,t-1+k/n}^2\:\:\:\forall i=1,...,M\:,\:~~~~~~~\hat{\sigma}_{f}^2=\frac{1}{n-1}\sum_{k=1}^{n}f_{t-1+k/n}^2.
\end{equation}
Since by assumption residuals are i.i.d. and normal, the quantities $(n-1)\hat{\sigma}_{\epsilon}^2/\sigma_{\epsilon}^2$ and $(n-1)\hat{\sigma}_{f}^2/\sigma_{f}^2$
follow a chi-squared distribution $\chi^2_{n-1}$
with $n-1$ degrees of freedom, and the $90\%$ confidence interval of $\hat{\sigma}_{\epsilon,f}^2$ is
\begin{equation}\label{fluctuation}
\delta\hat{\sigma}_{\epsilon,f}^2\equiv\frac{\sigma_{\epsilon,f}^2}{n-1}[(\chi^2_{n-1})^{-1}(0.95)-(\chi^2_{n-1})^{-1}(0.05)].
\end{equation}
We use this confidence interval as a measure of the fluctuations in risk estimation. Let us notice that it goes to zero when $n$ goes to infinity, as numerically confirmed in the simulation of the model. Fluctuations in the estimation of the covariance matrix in Eq. \ref{covendo} are obtained according to formulas in Appendix \ref{covarianceAppendix}, see Eqs. \ref{varendoApp}, \ref{covendoApp}, \ref{SigmaTheo}\footnote{From the point of view of VAR(1) process estimation, model variables $\lambda_{t-1},m_{t-1}$ in formulas in Appendix \ref{covarianceAppendix} correspond to parameters of the VAR(1) process, $\Phi_{t-1}$ in Equation \ref{fast_d}. Similarly to the variance of residuals $\hat{\sigma}_{\epsilon}^2$ and $\hat{\sigma}_{f}^2$, in the ideal process of estimating $\bm{\Phi}_{t-1}$, we obtain in turn the estimator $\widehat{\bm{\Phi}}_{t-1}$ within confidence intervals. For simplicity, we make mean field approximation by assuming a picked distribution on the real values.}.

Portfolio decisions are directly affected by fluctuations in risk estimation. Through the analytical mapping of Eq. \ref{dyn_s} and by using $\Sigma_\epsilon\equiv n\sigma_\epsilon^2$ and $\Sigma_f\equiv n\sigma_f^2$, we obtain the range in which leverage varies depending on $\delta\hat{\sigma}_{\epsilon,f}^2$. That is, the $90\%$ confidence interval of leverage, $\delta\lambda$, with respect to the skeleton dynamics when $n$ is finite\footnote{$\delta\lambda$ is obtained by differentiating Eq. \ref{dyn_s} with respect to $\lambda_t$ and $\sigma_{\epsilon,f}^2$.}.

Figure \ref{fluctuations_fig} shows the deterministic skeleton and the fluctuations around it when $n$ is finite. When $n$ is large, the dynamical evolution of the system is very close to its skeleton. By decreasing the value for $n$, fluctuations become more and more important. Empirically we observe that $\delta\lambda$ is inversely proportional to $n$. However, at least for the 2-period cycles, properties of the dynamical model are conserved when $n$ is of order $\mathcal{O}(10^3)$ or larger.

Notice that the approach presented here can not be applied to the chaotic dynamics. In dynamical systems theory, when chaos occurs, it is not possible to describe the system through the linearization around an equilibrium because there is not anymore a generalized equilibrium point (such as a stable fixed point or a periodic orbit), see \cite{eckmann1985ergodic}. Hence, this analytical argument based on the perturbation of a generalized equilibrium can not be applied when the deterministic skeleton is chaotic.

\subsection{Numerical results}\label{1D}

In this Subsection we study numerically the model for finite $n$. For tractability, we focus on the reduced version of the original model by considering one financial institution investing in one risky asset. In the reduced model, we lose the aspects related to the diversification of the portfolio. In turn, the comparison of numerical simulations with the theoretical results of the previous Subsection is easily obtained, as it will be clear below. Finally, this approach via the reduced model focuses more clearly on the role of $n$ in systemic stability of the financial system. In the numerical simulations, we assume that the bank correctly perceives the autoregressive dynamics of the asset price and the variance $\hat{\sigma}_t^2$ is obtained by estimating the AR(1) process on observed returns, see Eq. \ref{n1}. We consider for simplicity the case of large memory in risk expectations ($\omega=0.4$) in such a way that the deterministic skeleton is a fixed point equilibrium. The case of 2-period cycles is equivalent from the point of view of the analytical approximations. 
\begin{figure}[h]
\centering
{
	\includegraphics[width=0.88\textwidth]{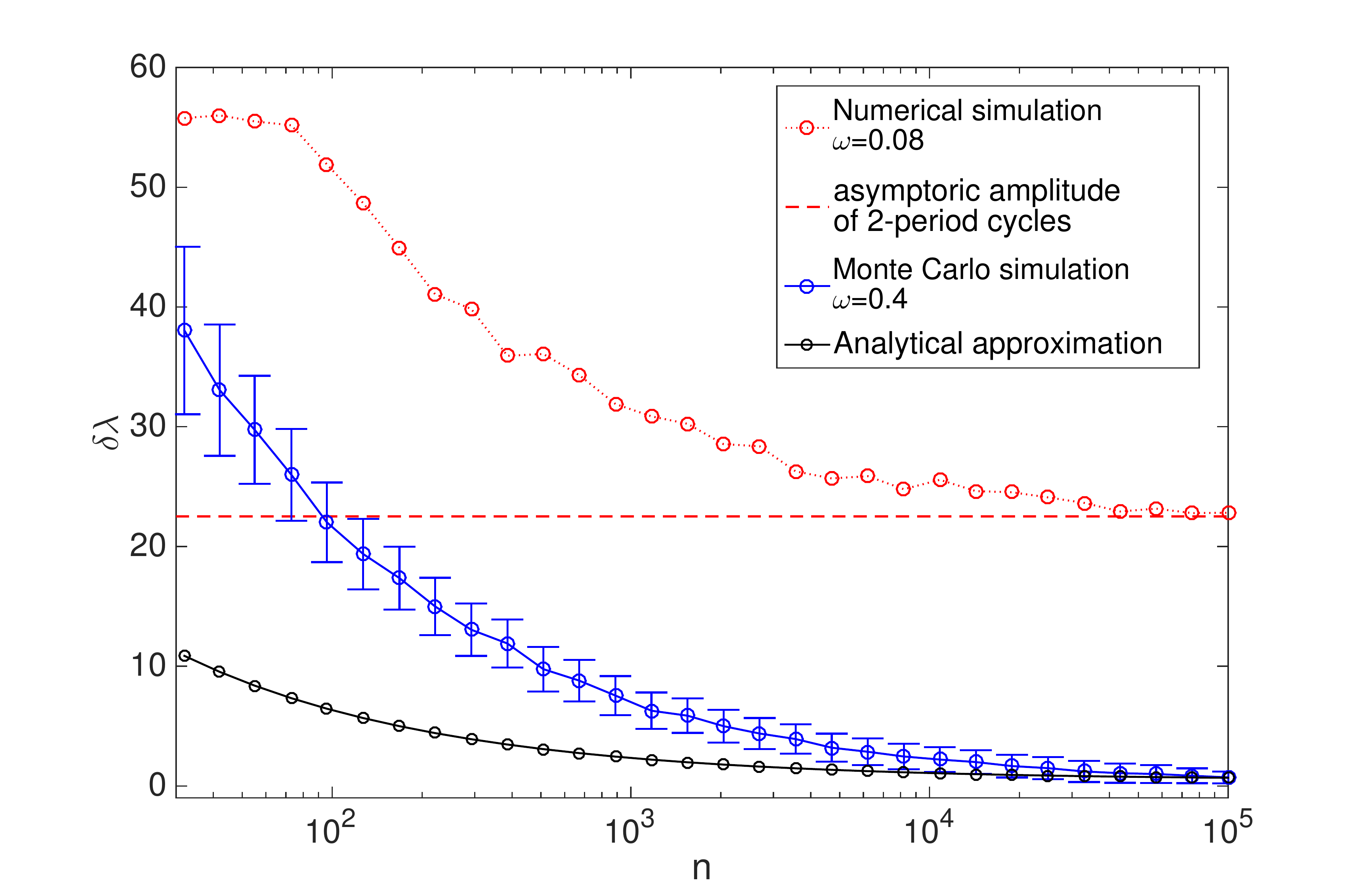}}
\caption{\footnotesize{$90\%$ confidence interval  $\delta\lambda$ , see Eq. \ref{lflun}, as a function of $n$ for two different values of $\omega$: $\omega=0.4$ (blue dots) and $\omega=0.08$ (red dots). Theoretical prevision for $\omega=0.4$ obtained via Eq. \ref{fluctuation} in the case $N=M=1$ is represented by the black dots. In the Monte Carlo simulations of the reduced model, for each value of $n$ we average over 50 seeds to obtain the considered value for $\hat{\delta\lambda}$ and the error bar represents the standard deviation of the data. The other model parameters are: $\gamma=100$, $\alpha=1.64$, $\Sigma_\epsilon=0.05$, $T=2000$.}}
\label{approx}
\end{figure}

In Figure \ref{approx} we show the fluctuations of the financial leverage $\delta\lambda$ as a consequence of the fluctuations in the estimator of risk. We compare the analytical approximation (black dots) with Monte Carlo simulations (blue dots). The analytical approximation is obtained via Eq. \ref{fluctuation} with $M=1$ risky investment. Numerically, we obtain $\hat{\delta\lambda}$ in the following way. We simulate the reduced version of the slow-fast random dynamical system. Then, we collect the data after the initial transient is passed. Hence we obtain the empirical probability distribution $\hat{F}_\lambda$ associated with a simulated path $\{\hat{\lambda}_t\}_{t=0,...,T}$. Coherently with $\delta\lambda$ obtained via the analytical approximation, we determine $\hat{\delta\lambda}$ as the $90\%$ confidence interval
\begin{equation}\label{lflun}
\hat{\delta\lambda}=\hat{F}_\lambda^{-1}(0.95)-\hat{F}_\lambda^{-1}(0.05).
\end{equation}
The theoretical prediction and the numerical results agree when $n\geq10^4$.

\paragraph{{\bfseries Discussion: trading costs}}
In the recent literature on leverage cycles, the possibility of two different timescales for portfolio decisions and leverage targeting is not analyzed. We strongly believe that this aspect deserves more attention since it is directly related to the discussion about the role of taxation in financial systemic risk \cite{matheson2012security,masciandaro2013financial}. 
We can interpret $n$ as an indirect measure of financial transaction taxes (e.g. the Tobin tax) and other trading frictions. Indeed, $n$ is related to the number of trading operations in the market, because it is equal to the number of rebalancing operations by a bank to have the balance sheet close to the desired capital structure. Clearly high transaction taxes decrease the number of operations for a financial institution. 

In our analysis, we study how the amplitude of leverage cycles depends on $n$ in the reduced model. In Figure \ref{approx}, we show $\delta\lambda$ as a function of $n$. The plot shows the results of Monte Carlo simulations and each point is obtained by averaging over $50$ realizations of the system dynamics with $T=2000$. We show the obtained results for two different values of $\omega$. The case $\omega=0.4$ is associated with the fixed point equilibrium in the deterministic skeleton, while the case $\omega=0.08$ corresponds to the 2-period cycles. When $n\gg1$ we recover the deterministic skeleton as shown in the previous Section. In this limit, $\delta\lambda$ tends to zero in the case of fixed point equilibrium because oscillations tend to disappear. When the deterministic skeleton corresponds to the 2-period cycles, the asymptotic value of $\delta\lambda$ corresponds to the amplitude of the bifurcated 2-period orbit. When $n\gg1$, both simulated and theoretical amplitude of the leverage cycles tend to coincide, that is a further proof of the consistency of the analytical approximation introduced previously.

In both cases, we can notice that the amplitude of leverage cycles is a decreasing function of $n$. The intuition is that when $n$ is small, leverage oscillations tend to correlate more strongly with the price movements. Hence, increments in target leverage reflects the increasing of asset size and at the same time perception of low risks.  However, since the bank is not marked-to-market but it has increased considerably its target leverage, when an exogenous shock for the price occurs, a panic-induced fall in financial leverage follows. On the contrary, when $n$ is large and the bank is marked-to-market in the capital structure, the financial system tends to be close to its deterministic skeleton. In the case of fixed point equilibrium, the leverage cycles tend to disappear because of exact risk previsions, at least in the long run. 

Also when the adopted memory in risk expectations is small and the deterministic skeleton corresponds to the 2-period orbit, the amplitude of the leverage cycles is smaller for larger $n$. Furthermore, in this case the cyclical structure of the dynamics can be easily recognized, e.g. in the autocorrelation functions of the realized variance. This information can be usefully adopt to improve risk estimations.

Our results suggest that allowing mark-to-market in the capital structure by removing trading frictions tends to reduce the amplitude of leverage cycles and as a consequence the financial system is closer to the fixed point equilibrium.

\section{Conclusions}\label{conclusions}
In this paper, we study the implications of backward-looking expectations of risk on portfolio decisions and as a consequence on the stability properties of the financial system. To this end we develop a model which becomes a slow-fast random dynamical system describing the evolution of a bipartite financial network of risky investments and bank assets. The main characteristics of this stylized financial market are: (i) financial institutions having capital requirements in the form of the VaR constraint and following standard mark-to-market and risk management rules; (ii) the presence of asset illiquidity; (iii) indirect contagion of risk mediated by the overlapping of portfolios; (iv) backward-looking expectations via statistical models of past observations of prices.

In the asymptotic deterministic limit, we are able to study analytically the fixed point equilibrium of the financial system and how the breaking of systemic stability occurs. The main result of this paper is the analytical classification of the possible dynamical outcomes for the considered financial system and its relation with market conditions that are represented by the memory of expectations $\omega$, the probability of Value at Risk defined by $\alpha$, and the cost of diversification $c$. We show how the breaking of the fixed point equilibrium for the financial system occurs via a period-doubling bifurcation which determines the appearing of leverage cycles. Furthermore, we show that the dynamics of the financial system is chaotic in certain limits and, at the best of our knowledge, this paper represents the first analytical proof of the appearing of chaotic dynamics in this context. Some recent literature, see for example \cite{aymanns2015dynamics}, has argued the presence of chaotic attractors in models of systemic risk in financial systems by means of numerical arguments. Then, \cite{douady2012financial} proposed a new instability indicator whose goal is capturing the chaotic dynamics of cash flows among financial institutions during turmoil periods, and \cite{castellacci2014modeling} have modeled financial contagion in the Eurozone crisis with a similar aim.

Finally, we study the dynamical outcomes of the model in order to answer the following questions, within this model:
\begin{enumerate}
\item \emph{How do the system's stability properties depend on expectations of risk formed by banks?} Our answer is that, all else being equal, the larger is the memory $\omega$ in the process of expectations formation, the more stable is the financial system dynamics.
\item \emph{How important is the constraint on financial leverage?} In our analysis we show that, for any value of $n$, there exists a tipping point for $\alpha$ which defines a `transition' from the fixed point equilibrium to the cyclical evolution. Hence, a more stringent regulation for the financial leverage is always stabilizing for the market.
\item \emph{What are the consequences of introducing new financial instruments?} We find that a market with a larger number of asset investments requires larger memory in forming risk expectations to be in dynamical equilibrium. Indeed, a larger number $M$ of financial instruments decreases the risk perception by banks because of larger portfolio diversification, that is followed by an increase of the financial leverage. The combined effects of the overlap of banks' portfolios with the price impact of the leverage targeting make the system more unstable.

\item \emph{What is the role of market frictions from a systemic risk point of view?} Decreasing the cost of diversification, here represented by $c$, may appear positive from a microscopic point of view but may lead to increase coordination of feedback effects due to similarity of banks portfolios, triggering a transition from a stable dynamics to the unstable one. On the contrary, our analysis suggests that decreasing transaction costs and removing all trading frictions may induce financial investors to adopt the strategy of being marked-to-market in their capital structure, \ie large values for the control parameter $n$. Within our model, this represents the control strategy of the balance sheet which has the consequence of reducing the amplitude of the cycles of leverage.
\end{enumerate}
A challenging issue, which will lead to further developments of this work, refers to relax the homogeneity assumptions regarding investment assets and/or financial institutions and the introduction of different time horizons in the process of risk expectations formation.

\section*{Acknowledgements}
We acknowledge Fulvio Corsi for the useful discussions and we are grateful for suggestions we have received from participants to the $18^{th}$ Workshop in Quantitative Finance in Milan and the $29^{th}$ Annual Conference of the European Association for Evolutionary Political Economy in Budapest. Finally, we are grateful to UniCredit Bank R\&D group for financial support through the "Dynamics and Information Theory Research Institute" at the Scuola Normale Superiore.

\appendix
\section{Covariance matrix of VAR(1) process at the slow time scale}\label{covarianceAppendix}
In this Section we show how to find the maximum likelihood estimators of the (diversifiable) variance and covariance of the VAR(1) process in Eq. \ref{slow_d} and how to compute analytically the covariance matrix in Eq. \ref{covendo}. However, before considering the multivariate VAR(1) process, let us focus on the univariate AR(1) process of the reduced model for which the computation of the variance represents a simpler problem which could shed light on the subsequent multivariate case.


\paragraph{{\bf Variance of the AR(1) process at the slow time scale}} Let us consider the process AR(1) in Eq. \ref{n1}, \ie
\be\label{eqA11}
r_s=\epsilon_s+\phi\:r_{s-1/n},\:\:\:s=t-1+k/n,\:k=1,2,...,n
\ee
where $\epsilon_s\sim\mathcal{N}(0,\sigma_\epsilon^2)$ $\forall s$, $|\phi|<1$ for the assumption of covariance stationarity and starting point $r_{t-1}$. The variance of the variable $r_s$ is $Var[r_s]=\E[r_s^2]=\frac{\sigma_\epsilon^2}{1-\phi^2}$, see \cite{tsay2005analysis}. The variance of the process aggregated between $t-1$ and $t$, \ie $Var[\sum_{k=1}^nr_{t-1+k/n}]$, is
\be\label{eqA12}
\E\left[\left(\sum_{k=1}^nr_{t-1+k/n}\right)^2\right]=n\E[r_s^2]+2\left(n\sum_{k=1}^n\E[r_sr_{s-k/n}]-\sum_{k=1}^nk\E[r_sr_{s-k/n}]\right)
\ee
where we use the assumption of covariance stationarity between $t-1$ and $t$. It is $\E[r_sr_{s-k/n}]=\phi^k\E[r_s^2]$, $\forall s=t-1+k/n,\:\:k=1,2,...,n$, by applying recursively Eq. \ref{eqA11}. By exploiting this result, it is
\be\label{eqA13}
\sum_{k=1}^n\E[r_sr_{s-k/n}]=\E[r_s^2]\sum_{k=1}^n\phi^k=\E[r_s^2]\frac{\phi(1-\phi^n)}{1-\phi},
\ee
\be\label{eqA14}
\sum_{k=1}^nk\E[r_sr_{s-k/n}]=\E[r_s^2]\sum_{k=1}^nk\phi^k=\E[r_s^2]\frac{(n\phi-n-1)\phi^{n+1}+\phi}{(1-\phi)^2}.
\ee
By substituting Eqs. \ref{eqA13} and \ref{eqA14} in Eq. \ref{eqA12}, we obtain the expression for the variance in Eq. \ref{n1}. In the limit $n\rightarrow\infty$, it is $\phi^n\rightarrow 0$ and $n\phi^n\rightarrow 0$ because of $|\phi|<1$. Then, the formula for the variance in Eq. \ref{MAPPAVERA}.

\paragraph{{\bf Multivariate case: VAR(1)}}
Let us focus on the VAR(1) process described in Eq. \ref{fast_d} where the endogenous component follows the VAR(1) process of Eq. \ref{endo2}. Let us assume $n\gg1$ for analytical tractability. At the time scale of the portfolio rebalancing, \ie $\frac{1}{n}$, the variance and the covariance associated with the endogenous component can be computed analytically (see \cite{corsi2013when}). When $s=t-1+k/n,\:\:k=1,2,...,n$, it is
\begin{equation}\label{varendoApp}
\begin{split}
Var[e_{i,s}] &=-\frac{\tilde{\lambda}^2}{(\gamma^2-\tilde{\lambda}^2)(m^2(\gamma^2(M-1)^2-\tilde{\lambda}^2)+2-\tilde{\lambda}^2mM-\tilde{\lambda}^2M^2)}\times\\
\times&\left(m^2(\sigma_\epsilon^2(\tilde{\lambda}^2-\gamma^2(M-1))+\sigma_f^2(\tilde{\lambda}^2-\gamma^2(M-1)^2))+\right.\\
&+2m(M(\sigma_\epsilon^2(\gamma^2-\tilde{\lambda}^2)-\tilde{\lambda}^2\sigma_f^2)-\gamma^2\sigma_\epsilon^2)+\\
&\left.+M(M(\sigma_\epsilon^2(\tilde{\lambda}^2-\gamma^2)+\tilde{\lambda}^2\sigma_f^2)+\gamma^2\sigma_\epsilon^2)\right),
\end{split}
\end{equation}
\begin{equation}\label{covendoApp}
\begin{split}
Cov[e_{i,s},e_{j,s}] &=-\frac{\tilde{\lambda}^2}{(\gamma^2-\tilde{\lambda}^2)(m^2(\gamma^2(M-1)^2-\tilde{\lambda}^2)+2-\tilde{\lambda}^2mM-\tilde{\lambda}^2M^2)}\times\\
\times&\left(m^2(\sigma_f^2(\tilde{\lambda}^2-\gamma^2(M-1)^2)-\gamma^2(M-2)\sigma_\epsilon^2)+\right.\\
&\left.-2m(\tilde{\lambda}^2M\sigma_f^2+\gamma^2\sigma_\epsilon^2)+M(\tilde{\lambda}^2M\sigma_f^2+\gamma^2\sigma_\epsilon^2)\right),
\end{split}
\end{equation}
where we have defined the excess leverage as $\tilde{\lambda}\equiv\lambda-1$ and for notational simplicity we do not label $\lambda$ and $m$ with the time index $t-1$. Hence, at the time scale of the portfolio rebalancing the variance and the covariance of the returns, $r_{i,s}$\footnote{For simplicity, we are assuming that the asset returns are centered around the mean.}, are 
\begin{equation}\label{varrApp}
Var[r_{i,s}]=\sigma_\epsilon^2+\sigma_f^2+Var[e_{i,s}]\equiv\theta_0
\end{equation}
\begin{equation}\label{covrApp}
Cov[r_{i,s},r_{j,s}]=\sigma_f^2+Cov[e_{i,s},e_{j,s}]\equiv\psi_0,
\end{equation}
respectively. The variance and the covariance of the returns at the time scale of the portfolio decisions, \ie $R_{i,t}\equiv\sum_{k=1}^n r_{i,t-1+k/n}$, are
$$
Var[R_{i,t}]=\E[(R_{i,t})^2]=n\E[r_{i,s}^2]+2(n-1)\E[r_{i,s}r_{i,s-1/n}]+2(n-2)\E[r_{i,s}r_{i,s-2/n}]+...
$$
$$
Cov[R_{i,t},R_{j,t}]=\E[R_{i,t}R_{j,t}]=n\E[r_{i,s}r_{j,s}]+2(n-1)\E[r_{i,s}r_{j,s-1/n}]+2(n-2)\E[r_{i,s}r_{j,s-2/n}]+...,\:j\neq i
$$
because of the stationarity of the return dynamics between $t-1$ and $t$. By defining $\theta_k\equiv\E[r_{i,s}r_{i,s-k/n}]$ and $\psi_k\equiv\E[r_{i,s}r_{j,s-k/n}]$ with $j\neq i$, the previous formulas read as
\begin{equation}\label{varR}
Var[R_{i,t}]=n\: \theta_0+2n\sum_{k=1}^{n}\theta_k-2\sum_{k=1}^{n}k\:\theta_k
\end{equation}
\begin{equation}\label{covR}
Cov[R_{i,t},R_{j,t}]=n\: \psi_0+2n\sum_{k=1}^{n}\psi_k-2\sum_{k=1}^{n}k\:\psi_k.
\end{equation}

In a similar fashion of the univariate case, by applying recursively the VAR(1) process in Eq. \ref{fast_d} and by taking expectation, we obtain two systems of equations whose solutions are the analytical expressions of the terms $\sum_{k=1}^{n}\theta_k$, $\sum_{k=1}^{n}k\:\theta_k$, $\sum_{k=1}^{n}\psi_k$ and $\sum_{k=1}^{n}k\:\psi_k$ in Eqs. \ref{varR} and \ref{covR}.

For notational simplicity, let us define
\be\label{defVarphi}
\varphi\equiv\frac{\lambda-1}{\gamma\:m}
\ee
\be\label{defBeta}
\beta\equiv\frac{\lambda-1}{\gamma\:m}\frac{m-1}{M-1}
\ee
such that the matrix of autoregressive coefficients in Eq. \ref{fast_d} reads as $\bm{\Phi}=(\varphi-\beta)\mathbb{I}+\beta\bm{1}$ where $\mathbb{I}$ is the identity matrix and $\bm{1}$ is the matrix whose entries are equal to one.

It is possible to verify that $\Theta_1\equiv\sum_{k=1}^{n}\theta_k$ and $\Psi_1\equiv\sum_{k=1}^{n}\psi_k$ are the solutions of the following linear system of equations,
\begin{equation}\label{lsGeoSum}
\begin{cases}
(1-\varphi)\Theta_1-\beta(M-1)\Psi_1 &= \varphi\:\theta_0+\beta(M-1)\:\psi_0 \\
-\beta\:\Theta_1+(1-(\varphi+\beta(M-2)))\Psi_1 &= \beta\:\theta_0+(\varphi+\beta(M-2))\:\psi_0,
\end{cases}
\end{equation}
where we have assumed that $\theta_{n}\ll1$ and $\psi_{n}\ll1$. It holds when $n\gg1$, see \cite{tsay2005analysis}.

Similarly, $\Theta_2\equiv\sum_{k=1}^{n}k\:\theta_k$ and $\Psi_2\equiv\sum_{k=1}^{n}k\:\psi_q$ are the solutions of the following linear system of equations,
\begin{equation}\label{lsGeoSum2}
\begin{cases}
(1-\varphi)\Theta_2-\beta(M-1)\Psi_2 &= \varphi\:\Theta_1+\beta(M-1)\:\Psi_1 \\
-\beta\:\Theta_2+(1-(\varphi+\beta(M-2)))\Psi_2 &= \beta\:\Theta_1+(\varphi+\beta(M-2))\:\Psi_1,
\end{cases}
\end{equation}
where we have assumed $\theta_{n}\ll1$, $n\:\theta_{n}\ll1$, $\psi_{n}\ll1$ and $n\:\psi_{n}\ll1$ when $n\gg1$.

According to our assumptions about statistical equivalence of risky investments, the covariance matrix is a matrix whose diagonal entries are equal to each other, \ie $\bar{\Sigma}_d+\bar{\Sigma}_u\equiv Var[R_{i,t}]\:\forall i$, and the same for the off-diagonal ones, \ie $\bar{\Sigma}_u\equiv Cov[R_{i,t},R_{j,t}]\:\forall i\neq j$. Hence, for $n\gg1$ the covariance matrix of the returns in Eq. \ref{fast_d} at the time scale of the portfolio decisions is
\be\label{covMatrixExplicit}
\bar{\bm{\Sigma}}=\bar{\Sigma}_d\mathbb{I}+\bar{\Sigma}_u\bm{1}
\ee
with
\be\label{SigmaTheo}
\begin{cases}
\bar{\Sigma}_d=n\left((\theta_0-\psi_0)+2(\Theta_1-\Psi_1)-\frac{2}{n}(\Theta_2-\Psi_2)\right)\\
\bar{\Sigma}_u=n\left(\psi_0+2\Psi_1-\frac{2}{n}\Psi_2\right)
\end{cases}
\ee
and by substituting the formulas of Eqs. \ref{varendoApp}, \ref{covendoApp}, \ref{varrApp} and \ref{covrApp} in Eq. \ref{SigmaTheo}, the covariance matrix is obtained explicitly. 

\paragraph{{\bf Maximum likelihood estimation}}
In the ideal procedure of estimating the VAR(1) process in Eq. \ref{fast_d}, \ie
\begin{equation}\label{VAR1}
\bm{r}_s=\bm{\Phi}\:\bm{r}_{s-1/n}+\bm{\varepsilon}_s, \:\:s=t-1+k/n,\:\:k=1,2,...,n,
\end{equation}
we can obtain the maximum likelihood estimators of $\bar{\Sigma}_d$ and $\bar{\Sigma}_u$, \ie $\widehat{\Sigma}_d$ and $\widehat{\Sigma}_u$.

In Eq. \ref{VAR1}, $\bm{\varepsilon}_s\sim\mathcal{N}(0,\bm{\Sigma}_\varepsilon)$ $\forall s=t-1+k/n,\:k=1,...,n$ where $\bm{\Sigma}_\varepsilon=\sigma_\epsilon^2\mathbb{I}+\sigma_f^2\bm{1}$ with $\sigma_\epsilon,\sigma_f>0$ and $\bm{\Phi}=(\varphi-\beta)\mathbb{I}+\beta\bm{1}$ with $\varphi$ and $\beta$ defined in Eqs. \ref{defVarphi} and \ref{defBeta}.

Given the observations between $t-1$ and $t$, the likelihood of the VAR(1) process is
$$
\P[\bm{r}_{t-1+1/n},\bm{r}_{t-1+2/n},...,\bm{r}_{t}|\bm{r}_{t-1}]=\prod_{k=1}^n\P[\bm{r}_{t-1+k/n}|\bm{r}_{t-1+(k-1)/n}]=\prod_{k=1}^n\mathcal{N}(\bm{\Phi}\bm{r}_{t-1+(k-1)/n},\:\bm{\Sigma}_\varepsilon)
$$
because the conditional distribution of the returns is multivariate Gaussian. Hence, the log-likelihood is
\be\label{logLikelihoodVAR1}
L(\bm{\Phi},\bm{\Sigma}_\varepsilon)=-\frac{n}{2}\log|\bm{\Sigma}_\varepsilon|-\frac{1}{2}\sum_{k=1}^n(\bm{r}_{t-1+k/n}-\bm{\Phi}\:\bm{r}_{t-1+(k-1)/n})^\intercal\bm{\Sigma}_\varepsilon^{-1}(\bm{r}_{t-1+k/n}-\bm{\Phi}\:\bm{r}_{t-1+(k-1)/n})
\ee
and the maximum likelihood estimators of parameters, \ie $\hat{\varphi}$, $\hat{\beta}$, $\hat{\sigma}_\epsilon^2$ and $\hat{\sigma}_f^2$, can be obtained as solution of the following system of equations
{\large
$$
\begin{cases}
\frac{\partial L(\varphi,\beta,\sigma_\epsilon,\sigma_f)}{\partial\varphi} = 0\\
\frac{\partial L(\varphi,\beta,\sigma_\epsilon,\sigma_f)}{\partial\beta} = 0\\
\frac{\partial L(\varphi,\beta,\sigma_\epsilon,\sigma_f)}{\partial\sigma_\epsilon} = 0\\
\frac{\partial L(\varphi,\beta,\sigma_\epsilon,\sigma_f)}{\partial\sigma_f} = 0.
\end{cases}
$$}
Let us notice that the maximum likelihood estimators of parameters are only functions of the observed fast variables $\{\bm{r}_{t-1+k/n}\}_{k=0,1,2,...,n}$,
\ie $\hat{\varphi}\equiv \hat{\varphi}(\{\bm{r}_{t-1+k/n}\}_{k=0,1,2,...,n})$, $\hat{\beta}\equiv \hat{\beta}(\{\bm{r}_{t-1+k/n}\}_{k=0,1,2,...,n})$, $\hat{\sigma}_\epsilon\equiv \hat{\sigma}_\epsilon(\{\bm{r}_{t-1+k/n}\}_{k=0,1,2,...,n})$ and $\hat{\sigma}_f\equiv \hat{\sigma}_f(\{\bm{r}_{t-1+k/n}\}_{k=0,1,2,...,n})$.

Then, according to the following standard result of time series analysis, see \cite{tsay2005analysis},
$$
(\hat{\theta}_0-\hat{\psi}_0)\mathbb{I}+\hat{\psi}_0\bm{1}\equiv\sum_{k=0}^\infty\widehat{\bm{\Phi}}^k\widehat{\bm{\Sigma}}_\varepsilon(\widehat{\bm{\Phi}}^k)^\intercal
$$
where $\widehat{\bm{\Sigma}}_\varepsilon=\hat{\sigma}_\epsilon^2\mathbb{I}+\hat{\sigma}_f^2\bm{1}$ and $\widehat{\bm{\Phi}}=(\hat{\varphi}-\hat{\beta})\mathbb{I}+\hat{\beta}\bm{1}$, we are able to compute the maximum likelihood estimators of the variance and covariance at the fast time scale, \ie $\hat{\theta}_0\equiv\widehat{Var}[r_{i,s}]$ and $\hat{\psi}_0\equiv\widehat{Cov}[r_{i,s},r_{j,s}]$ with $j\neq i$. Hence, by computing $\hat{\Theta}_1\equiv\Theta_1(\hat{\theta}_0,\hat{\psi}_0,\hat{\varphi},\hat{\beta})$, $\hat{\Psi}_1\equiv\Psi_1(\hat{\theta}_0,\hat{\psi}_0,\hat{\varphi},\hat{\beta})$, $\hat{\Theta}_2\equiv\Theta_2(\hat{\theta}_0,\hat{\psi}_0,\hat{\varphi},\hat{\beta})$ and $\hat{\Psi}_2\equiv\Psi_2(\hat{\theta}_0,\hat{\psi}_0,\hat{\varphi},\hat{\beta})$, the maximum likelihood estimators of the (diversifiable) variance and covariance of the return dynamics in Eq. \ref{VAR1} aggregated at the slow time scale are
\be\label{SigmaHat}
\begin{cases}
\widehat{\Sigma}_d=n\left((\hat{\theta}_0-\hat{\psi}_0)+2(\hat{\Theta}_1-\hat{\Psi}_1)-\frac{2}{n}(\hat{\Theta}_2-\hat{\Psi}_2)\right)\\
\widehat{\Sigma}_u=n\left(\hat{\psi}_0+2\hat{\Psi}_1-\frac{2}{n}\hat{\Psi}_2\right).
\end{cases}
\ee

\paragraph{{\bf Asymptotic limit $n\rightarrow\infty$}}

In the limit $n\rightarrow\infty$ the last terms in Eq. \ref{SigmaTheo} are negligible. However, the terms $n\left((\theta_0-\psi_0)+2(\Theta_1-\Psi_1)\right)$ and $n\left(\psi_0+2\Psi_1\right)$ remain finite when $n$ goes to infinity because of finite $\Sigma_\epsilon=\lim_{n\rightarrow\infty}n\sigma_\epsilon^2$ and $\Sigma_f=\lim_{n\rightarrow\infty}n\sigma_f^2$.

By defining
{\SMALL
\be\label{Thetone0}
\Theta_0\equiv\Sigma_\epsilon+\Sigma_f-\frac{\tilde{\lambda}^2\left(m^2(\Sigma_\epsilon(\tilde{\lambda}^2-\gamma^2(M-1))+\Sigma_f(\tilde{\lambda}^2-\gamma^2(M-1)^2))+2m(M(\Sigma_\epsilon(\gamma^2-\tilde{\lambda}^2)-\tilde{\lambda}^2\Sigma_f)-\gamma^2\Sigma_\epsilon)+M(M(\Sigma_\epsilon(\tilde{\lambda}^2-\gamma^2)+\tilde{\lambda}^2\Sigma_f)+\gamma^2\Sigma_\epsilon)\right)}{(\gamma^2-\tilde{\lambda}^2)(m^2(\gamma^2(M-1)^2-\tilde{\lambda}^2)+2-\tilde{\lambda}^2mM-\tilde{\lambda}^2M^2)},
\ee
\be\label{Psone0}
\Psi_0\equiv\Sigma_f-\frac{\tilde{\lambda}^2\left(m^2(\Sigma_f(\tilde{\lambda}^2-\gamma^2(M-1)^2)-\gamma^2(M-2)\Sigma_\epsilon)-2m(\tilde{\lambda}^2M\Sigma_f+\gamma^2\Sigma_\epsilon)+M(\tilde{\lambda}^2M\Sigma_f+\gamma^2\Sigma_\epsilon)\right)}{(\gamma^2-\tilde{\lambda}^2)(m^2(\gamma^2(M-1)^2-\tilde{\lambda}^2)+2-\tilde{\lambda}^2mM-\tilde{\lambda}^2M^2)},
\ee
}
in the asymptotic limit $n\rightarrow\infty$, it is
\be\label{SigmaAsymptotic}
\begin{cases}
\widetilde{\Sigma}_d=\Theta_0-\Psi_0+2\frac{(\varphi\beta(M-2)+\varphi^2-\varphi-\beta^2(M-1))(\Theta_0-\Psi_0)+\beta\Theta_0+\beta(M-1)\Psi_0-(M-2)\Psi_0}{(1+\beta-\varphi)(\beta(M-1)+\varphi-1)}\\
\widetilde{\Sigma}_u=\Psi_0+2\frac{-\beta\Theta_0+(\varphi\beta(M-2)+\varphi^2-\varphi-\beta^2(M-1)+\varphi\beta(M-2))\Psi_0}{(1+\beta-\varphi)(\beta(M-1)+\varphi-1)}.
\end{cases}
\ee
Hence, by substituting Eqs. \ref{Thetone0} and \ref{Psone0} in Eq. \ref{SigmaAsymptotic} and since $\tilde{\lambda}\equiv\tilde{\lambda}_{t-1}$, $m_{t-1}\equiv m_{t-1}(\lambda_{t-1},\Sigma_{d,t-1},\Sigma_{u,t-1})$ according to Eq. \ref{mdependent}, we obtain the explicit analytical expressions for $\widetilde{\Sigma}_d$ and $\widetilde{\Sigma}_u$ in Eq. \ref{covendo}.

\section{Single time scale}\label{model1D-dt1}
Here we consider the case $n=1$ for the reduced model, \ie bank updates the risk expectation at the same frequency of portfolio rebalancing. When the time scales are the same, the process of expectation formation reduces to model the variance as in IGARCH-type models \cite{engle1986modelling},
\begin{equation}\label{igarch}
\sigma_t^2=\omega\sigma_{t-1}^2+(1-\omega)r_{t}^2
\end{equation}
with memory parameter $\omega$.

This approach is similar to the RiskMetrics one, see \cite{longerstaey1996riskmetricstm}, and, as highlighted by \cite{bauwens2006multivariate}, despite the simplicity, this kind of model is usually adopted by practitioners with a decay factor $\omega$ equal to $0.94$ for daily data and $0.97$ for monthly data. 

The model specified by Eqs. \ref{n1} with Eq. \ref{igarch} as process of expectation formation is close to the one presented in \cite{aymanns2015dynamics} which differs for the price dynamics. Aymanns and Farmer assume that the equity value is fixed and there is only one investor in the market, thus it is $A_t\equiv p_t=\lambda_tE$ with $p_t$ the price of the traded risky investment\footnote{Here, the number of shares is normalized to one.}. By assuming further the return as the log-difference $r_t=\log\frac{p_t}{p_{t-1}}$ and by substituting the price with $\lambda_t E$, a two-dimensional dynamical system is obtained. With this further assumption, the two models coincide.

\begin{figure}[t]
\centering
	\includegraphics[width=1\textwidth]{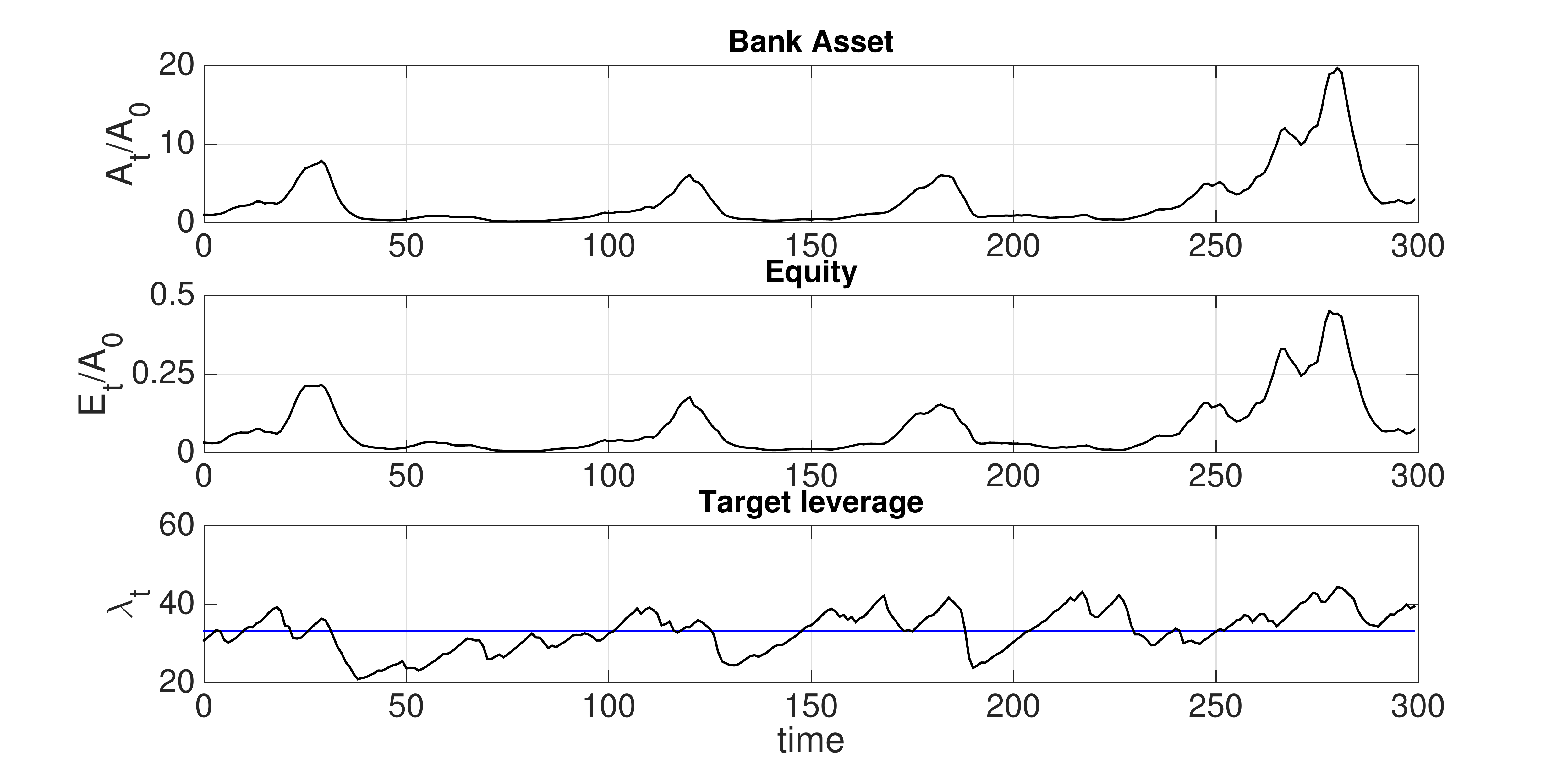}
\caption{\footnotesize{Numerical simulation of reduced model with $n=1$. Top panel: evolution of asset size normalized by the initial value $A_0$. Middle panel: evolution of equity. Bottom panel: dynamics for the financial leverage. We simulate the model in a time window $T=300$. The other model parameters are: $\gamma=40$, $\alpha=1.64$, $\mu-r_L=0.08$, $\sigma_\epsilon=0.04$, $\omega=0.9$.}}
\label{one_time_simu}
\end{figure}
Figure \ref{one_time_simu} shows the simulated dynamics of bank asset, equity and financial leverage of the reduced model with $n=1$. We notice that fluctuations in the value of target leverage are strongly correlated with fluctuations of the equity value. The changes in the equity value reflects the price movements. Asset size evolves in such a way to have financial leverage equal to its targeting value. Hence, boom and bust of the price are driven by leverage cycles which in turn are determined by risk perceptions.

As indicator of dynamical instabilities of the financial system we consider an adimensional measure of the amplitude of leverage cycles, specifically the standard deviation $\sigma_\lambda$ of target leverage in model simulations divided by its mean $\mu_\lambda$.

\begin{figure}[t]
\centering
{
	\includegraphics[width=0.65\textwidth]{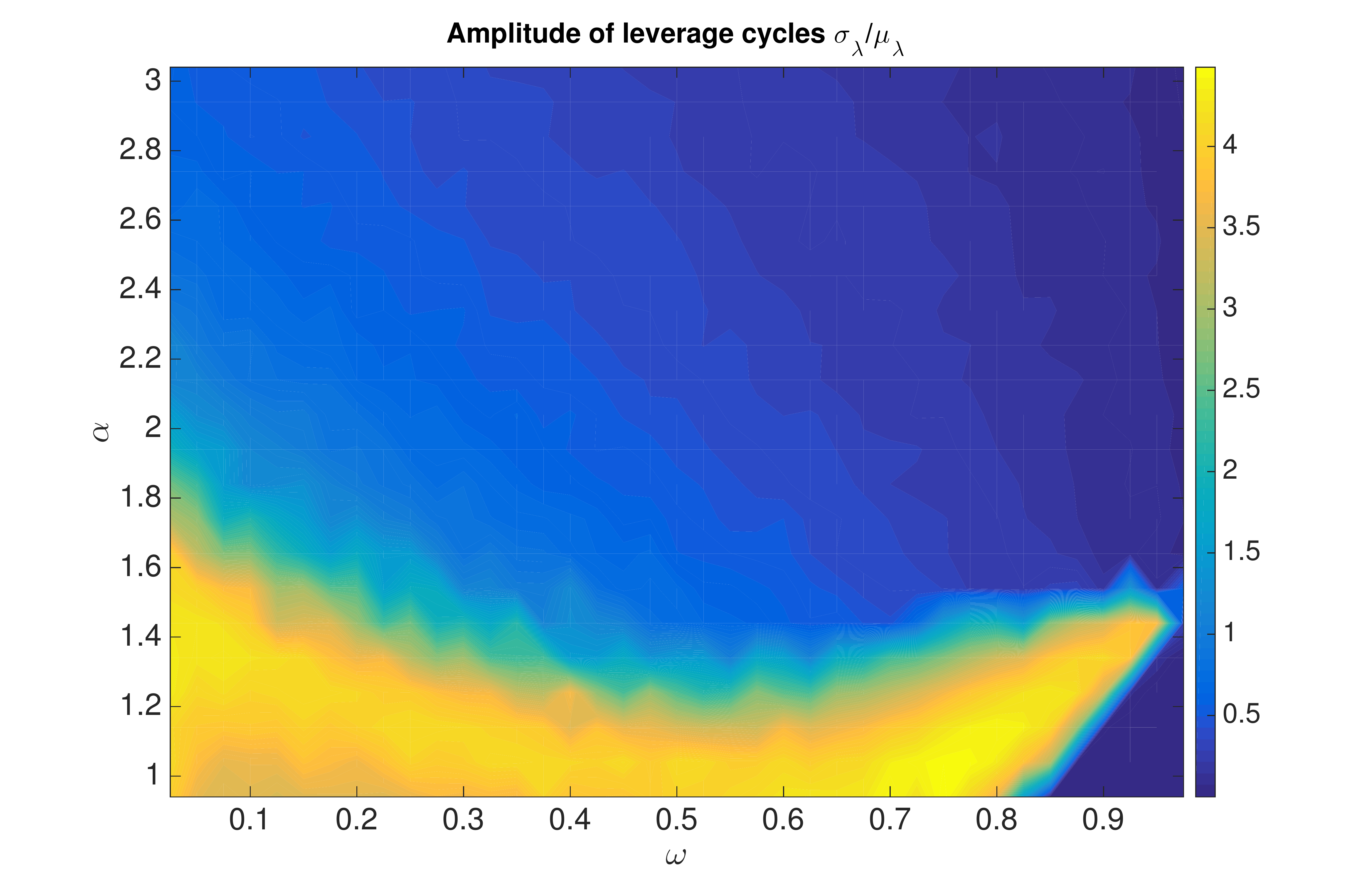}}
\caption{\footnotesize{The amplitude of leverage cycles  in relation with $\alpha$ and $\omega$. As measure of the amplitude of leverage cycles we use the standard deviation of data, $\sigma_\lambda$, divided by the mean $\mu_\lambda$. We run the reduced model for $T=1000$ with $\sigma_\epsilon^2=0.05$ and $\gamma=40$ by varying $\alpha\in[1,3]$ and $\omega\in(0,1)$. For each point $(\omega,\alpha)$ in the plot, we average over 50 seeds to obtain the considered value for $\frac{\sigma_\lambda}{\mu_\lambda}$.}}
\label{sigma_leva}
\end{figure}
In Figure \ref{sigma_leva} we show the normalized amplitude of leverage oscillations as a function of the VaR parameter $\alpha$ and of the memory parameter $\omega$. As expected, stringent capital constraints (large $\alpha$) and large memory in forecasting risk (large $\omega$) stabilize the dynamics of the financial system. Therefore, increasing $\alpha$ and $\omega$ is followed by smaller leverage oscillations and as a consequence the dynamics of financial system is close to the fixed point equilibrium. The parameter space for $\alpha$ and $\omega$ can be approximately divided in two regions. A region of instability for small value of $\alpha$ where strong oscillations occur independently on the memory of risk expectations and a relatively stable region for large $\alpha$ in which the coefficient $\frac{\sigma_\lambda}{\mu_\lambda}$ is smaller. In this region, larger memory tends to stabilize the system. Finally, let us notice that in the unstable region oscillations in financial leverage disappear for $\omega\rightarrow1^{-}$. We do not fully understand why we observe this but we  suppose the reason being the very high inertia in risk expectations which turns off the leverage oscillations.


\begin{thebibliography}{}
\bibitem[Adrian and Shin, 2010]{adrian2010liquidity}{Adrian, T., \& Shin, H. S. (2010). Liquidity and leverage. \emph{Journal of financial intermediation}, 19(3), 418-437.}
\bibitem[Adrian and Shin, 2014]{adrian2014procyclical}{Adrian, T., \& Shin, H. S. (2013). Procyclical leverage and value-at-risk. \emph{The Review of Financial Studies}, 27(2), 373-403.}
\bibitem[Aymanns et al., 2016]{aymanns2016taming}{Aymanns, C., Caccioli, F., Farmer, J. D., \& Tan, V. W. (2016). Taming the Basel leverage cycle. \emph{Journal of financial stability}, 27, 263-277.}
\bibitem[Aymanns and Farmer, 2015]{aymanns2015dynamics}{Aymanns, C., \& Farmer, J. D. (2015). The dynamics of the leverage cycle. \emph{Journal of Economic Dynamics and Control}, 50, 155-179.}
\bibitem[Bao et al., 2013]{bao2013learning}{Bao, T., Duffy, J., \& Hommes, C. (2013). Learning, forecasting and optimizing: An experimental study. \emph{European Economic Review}, 61, 186-204.}
\bibitem[Bauwens et al., 2006]{bauwens2006multivariate}{Bauwens, L., Laurent, S., \& Rombouts, J. V. (2006). Multivariate GARCH models: a survey. \emph{Journal of applied econometrics}, 21(1), 79-109.}
\bibitem[Bhattacharya and Majumdar, 2003]{bhattacharya2003random}{Bhattacharya, R., \& Majumdar, M. (2003). Random dynamical systems: a review. \emph{Economic Theory}, 23(1), 13-38.}
\bibitem[Brunnermeier and Pedersen, 2009]{brunnermeier2009market}{Brunnermeier, M. K., \& Pedersen, L. H. (2009). Market liquidity and funding liquidity. \emph{The Review of Financial Studies}, 22(6), 2201-2238.}
\bibitem[Caccioli et al., 2014]{caccioli2012stability}{Caccioli, F., Shrestha, M., Moore, C., \& Farmer, J. D. (2014). Stability analysis of financial contagion due to overlapping portfolios. \emph{Journal of Banking \& Finance}, 46, 233-245.}
\bibitem[Castellacci and Choi, 2014]{castellacci2014modeling}{Castellacci, G., \& Choi, Y. (2015). Modeling contagion in the Eurozone crisis via dynamical systems. \emph{Journal of Banking \& Finance}, 50, 400-410.}
\bibitem[Choi and Douady, 2012] {douady2012financial}{Choi, Y., \& Douady, R. (2012). Financial crisis dynamics: attempt to define a market instability indicator. \emph{Quantitative Finance}, 12(9), 1351-1365.}
\bibitem[Constantinides, 1986]{constantinides1986capital}{Constantinides, G. M. (1986). Capital market equilibrium with transaction costs. \emph{Journal of political Economy}, 94(4), 842-862.}
\bibitem[Cont and Wagalath, 2013]{cont2013running}{Cont, R., \& Wagalath, L. (2013). Running for the exit: distressed selling and endogenous correlation in financial markets. \emph{Mathematical Finance}, 23(4), 718-741.}
\bibitem[Corsi et al., 2016]{corsi2013when}{Corsi, F., Marmi, S., \& Lillo, F. (2016). When micro prudence increases macro risk: The destabilizing effects of financial innovation, leverage, and diversification. \emph{Operations Research}, 64(5), 1073-1088.}
\bibitem[Crawford, 1991]{crawford1991introduction}{Crawford, J. D. (1991). Introduction to bifurcation theory. \emph{Reviews of Modern Physics}, 63(4), 991.}
\bibitem[Danielsson et al., 2004]{danielsson2004impact}{Danielsson, J., Shin, H. S., \& Zigrand, J. P. (2004). The impact of risk regulation on price dynamics. \emph{Journal of Banking \& Finance}, 28(5), 1069-1087.}
\bibitem[Danielsson et al., 2012a]{danielsson2012endogenous}{Danielsson, J., Shin, H. S., \& Zigrand, J. P. (2012). Endogenous and systemic risk. In \emph{Quantifying systemic risk} (pp. 73-94). University of Chicago Press.}
\bibitem[Danielsson et al., 2012b]{danielsson2012endogenous_extreme}{Danielsson, J., Shin, H. S., \& Zigrand, J. P. (2012). Endogenous extreme events and the dual role of prices. \emph{Annu. Rev. Econ.}, 4(1), 111-129.}
\bibitem[Dolgopyat, 2004]{dolgopyat2004evolution}{Dolgopyat, D. (2004). Evolution of adiabatic invariants in stochastic averaging. \emph{Stochastics and Dynamics}, 4(02), 265-275.}
\bibitem[Dolgopyat, 2005]{dolgopyat2005introduction}{Dolgopyat, D. (2005). Introduction to averaging. (https://www.math.umd.edu/$\sim$dolgop/IANotes.pdf)}
\bibitem[Eckmann and Ruelle, 1985]{eckmann1985ergodic}{Eckmann, J. P., \& Ruelle, D. (1985). Ergodic theory of chaos and strange attractors. In \emph{The Theory of Chaotic Attractors} (pp. 273-312). Springer, New York, NY.}
\bibitem[Engle and Bollerslev, 1986]{engle1986modelling}{Engle, R. F., \& Bollerslev, T. (1986). Modelling the persistence of conditional variances. \emph{Econometric reviews}, 5(1), 1-50.}
\bibitem[Farmer et al., 2012]{farmer2012complex}{Farmer, J. D., Gallegati, M., Hommes, C., Kirman, A., Ormerod, P., Cincotti, S., Sanchez, A., \& Helbing, D. (2012). A complex systems approach to constructing better models for managing financial markets and the economy. \emph{The European Physical Journal - Special Topics}, 214(1), 295-324.}
\bibitem[Feigenbaum, 1978]{feigenbaum1978quantitative}{Feigenbaum, M. J. (1978). Quantitative universality for a class of nonlinear transformations. \emph{Journal of statistical physics}, 19(1), 25-52.}
\bibitem[Geanakoplos, 2010]{geanakoplos2010leverage}{Geanakoplos, J. (2010). The leverage cycle. \emph{NBER Macroeconomics Annual 2009}, 24(1), 1-66.}
\bibitem[Greenwood et al., 2015]{greenwood2015vulnerable}{Greenwood, R., Landier, A., \& Thesmar, D. (2015). Vulnerable banks. \emph{Journal of Financial Economics}, 115(3), 471-485.}
\bibitem[Halling et al., 2016]{halling2016leverage}{Halling, M., Yu, J., \& Zechner, J. (2016). Leverage dynamics over the business cycle. \emph{Journal of Financial Economics}, 122(1), 21-41.}
\bibitem[Heemeijer et al., 2009]{heemeijer2009price}{Heemeijer, P., Hommes, C., Sonnemans, J., \& Tuinstra, J. (2009). Price stability and volatility in markets with positive and negative expectations feedback: An experimental investigation. \emph{Journal of Economic dynamics and control}, 33(5), 1052-1072.}
\bibitem[Hommes, 1994]{hommes1994dynamics}{Hommes, C. H. (1994). Dynamics of the cobweb model with adaptive expectations and nonlinear supply and demand. \emph{Journal of Economic Behavior \& Organization}, 24(3), 315-335.}
\bibitem[Hommes, 2000]{hommes2000cobweb}{Hommes, C. (2000). Cobweb dynamics under bounded rationality. In \emph{Optimization, Dynamics, and Economic Analysis} (pp. 134-150). Physica, Heidelberg.}
\bibitem[Hommes, 2009]{hommes2009handbook}{Hommes, C. H., \& Wagener, F. (2009). Bounded rationality and learning in complex markets. \emph{Handbook of Economic Complexity}, 87-123.}
\bibitem[Hommes, 2013]{hommes2013behavioral}{Hommes, C. (2013). \emph{Behavioral rationality and heterogeneous expectations in complex economic systems}. Cambridge University Press.}
\bibitem[Hommes et al., 2007]{hommes2007learning}{Hommes, C., Sonnemans, J., Tuinstra, J., \& Van De Velden, H. (2007). Learning in cobweb experiments. \emph{Macroeconomic Dynamics}, 11(S1), 8-33.}
\bibitem[Kifer, 2014]{kifer2014nonconventional}{Kifer, Y. (2014). Nonconventional limit theorems in averaging. In \emph{Annales de l'Institut Henri Poincaré, Probabilités et Statistiques} (Vol. 50, No. 1, pp. 236-255). Institut Henri Poincaré.}
\bibitem[Kuehn, 2011]{kuehn2011mathematical}{Kuehn, C. (2011). A mathematical framework for critical transitions: Bifurcations, fast-slow systems and stochastic dynamics. \emph{Physica D: Nonlinear Phenomena}, 240(12), 1020-1035.}
\bibitem[Longerstaey and Spencer, 1996]{longerstaey1996riskmetricstm}{Longerstaey, J., \& Spencer, M. (1996). Riskmetricstm - technical document. \emph{Morgan Guaranty Trust Company of New York: New York.}}
\bibitem[Masciandaro and Passarelli, 2013]{masciandaro2013financial}{Masciandaro, D., \& Passarelli, F. (2013). Financial systemic risk: Taxation or regulation?. \emph{Journal of Banking \& Finance, 37(2), 587-596.}}
\bibitem[Matheson, 2012]{matheson2012security}{Matheson, T. (2012). Security transaction taxes: issues and evidence. \emph{International Tax and Public Finance}, 19(6), 884-912.}
\bibitem[May, 1976]{may1976simple}{May, R. M. (1976). Simple mathematical models with very complicated dynamics. \emph{Nature}, 261(5560), 459.}
\bibitem[Poledna et al., 2014]{poledna2014leverage}{Poledna, S., Thurner, S., Farmer, J. D., \& Geanakoplos, J. (2014). Leverage-induced systemic risk under Basle II and other credit risk policies. \emph{Journal of Banking \& Finance}, 42, 199-212.}
\bibitem[Tasca and Battiston, 2013]{tasca2013market}{Tasca, P., \& Battiston, S. (2016). Market procyclicality and systemic risk. \emph{Quantitative Finance}, 16(8), 1219-1235.}
\bibitem[Tsay, 2005]{tsay2005analysis}{Tsay, R. S. (2005). \emph{Analysis of financial time series} (Vol. 543). John Wiley \& Sons.}
\bibitem[Wolf et al., 1985]{wolf1985determining}{Wolf, A., Swift, J. B., Swinney, H. L., \& Vastano, J. A. (1985). Determining Lyapunov exponents from a time series. \emph{Physica D: Nonlinear Phenomena}, 16(3), 285-317.}



\end{thebibliography}
\end{document}